\documentclass[prd,twocolumn,secnumarabic,nobalancelastpage,amsmath,amssymb, nofootinbib, numerical]{revtex4-1}
\usepackage{graphics}      
\usepackage{graphicx}      
\usepackage{amsmath}
\usepackage{color}
\usepackage[normalem]{ulem}
\usepackage[letterpaper, left= 0.6in,right=0.6in,top=0.6in,bottom=0.6in, footskip=.25in]{geometry}

\begin{document}

\title{Coherent Detection of Ultra-weak Electromagnetic Fields}
\author{Zachary R. Bush\textsuperscript{1}, Simon Barke\textsuperscript{1}, Harold Hollis\textsuperscript{1}, Aaron D. Spector\textsuperscript{2}, Ayman Hallal\textsuperscript{1}, Giuseppe Messineo\textsuperscript{1}, D.B. Tanner \textsuperscript{1}, Guido Mueller\textsuperscript{1}}
\affiliation    {\textsuperscript{1}Department of Physics, University of Florida, PO Box 118440, Gainesville, Florida, 32611, USA \\
\textsuperscript{2}Deutsches Elektronen-Synchrotron (DESY), Notkestraße 85, D-22607 Hamburg, Germany}
\date{\today}

\begin{abstract}We explore the application of heterodyne interferometry for a weak-field coherent detection scheme. The methods detailed here will be used in ALPS II, an experiment designed to search for weakly-interacting, sub-eV particles. For ALPS II to reach its design sensitivity this detection system must be capable of accurately measuring fields with equivalent amplitudes on the order of 10$^{-5}$ photons per second or greater. We present initial results of an equivalent dark count rate on the order of $10^{-5}$ photons per second as well as successful generation and detection of a signal with a field strength equivalent to $10^{-2}$ photons per second.
\bigskip
 
 
\end{abstract}

\maketitle

\section{Introduction}
\subsection{Axions and Axion-Like Particles}
The Standard Model (SM) incorporates our current knowledge of subatomic particles as well as their interactions via three of the four fundamental forces of nature. The SM is not complete, however, as it does not contain gravity and does not explain certain observations. One notable unresolved issue is that of charge-conjugation parity symmetry (CP-symmetry) violation. The QCD Lagrangian includes terms capable of breaking CP-symmetry for the strong force. In contrast, experiments found that the strong forces respect CP-symmetry to a very high precision \cite{Neutron}.

The most prominent proposed solution,  introduced by Peccei and Quinn \cite{PhysRevLett.38.1440}, involves spontaneously breaking a global U(1) symmetry leading to a new particle, named the axion \cite{Weinberg, Wilczek}. Interactions with the QCD vacuum cause the axion to have a non-zero mass, $m_a$ \cite{PhysRevLett.38.1440}. While axions may interact with SM particles, the interactions can be weak. Most notably for experimental purposes, axion mixing with neutral pions leads to a characteristic two photon coupling, $g_{a \gamma \gamma}$ \cite{ParticleReview}. This, in turn, constrains the product of the axion mass and coupling such that these two parameters are dependent. Experimental and observational factors place the axion mass between 1 and 1000 $\mu$eV. The corresponding range for $g_{a \gamma \gamma}$ is $10^{-16}$ to $10^{-13}$ GeV$^{-1}$. 


While the QCD axion is confined to a specific band in the parameter space, it might just be a member of a larger class of axion-like particles, some with a stronger two-photon coupling \cite{Svrcek, ExperimentalSearch}. The interactions between these axions/axion-like particles and photons may possibly explain unanswered astronomical questions including TeV photon transparency in the Universe \cite{Meyer2013} and anomalous white dwarf cooling \cite{WhiteDwarf}. The intrinsic properties of axions and axion-like particles also make them prime candidates for cold dark matter. This theoretical motivation has led to the formulation of various experiments designed to detect axions and axion-like particles by utilizing their coupling to photons. 

Although axions can naturally decay into two observable photons, the rate at which this occurs is extremely low, making detection by observing this decay impossible. Axion search experiments therefore also rely on the inverse Primakoff or Sikivie effect in which a strong static magnetic field acts as a high density of virtual photons. This field stimulates the axion/axion-like particle to convert into a photon carrying the total energy of the axion/axion-like particle \cite{sikivie1983experimental, PhysRev.81.899}. A number of strategies have been employed by these experiments to search for axions from several sources. Haloscope experiments, such as ADMX, use resonant microwave cavities and strong superconducting magnets to search for axions comprising the Milky Way's cold dark-matter halo \cite{ADMXSuggested}. Helioscope experiments, such as CAST, look for relativistic axions originating from the Sun that convert into detectable X-rays as they pass through a supplied magnetic field \cite{PhysRevLett.94.121301}.  Differing from these types of axion searches that rely on astronomical sources, ``Light Shining through Walls'' (LSW) experiments attempt to generate and detect axions in the laboratory and therefore have the advantage of independence from models of the galactic halo and models of stellar evolution \cite{Bibber, Robilliard, GammeVSuggested, LIPPSSuggested, NewALPSResult, OSQARSuggested}.  

\subsection{ALPS II}
LSW experiments use the axion-photon coupling first to convert photons into axions under the presence of a strong magnetic field. These axions then pass through a light-tight barrier and enter another strong magnetic field where some are converted back into detectable photons. Energy is conserved in the process so that the regenerated photons  behind the wall have the same energy as those incident in front of it. The Any Light Particle Search (ALPS) is one such LSW experiment. The first generation of this experiment, ALPS I, set the most sensitive laboratory experimental limits of its time on the coupling of axions to two photons, $g_{a\gamma\gamma}$, for a wide range of axion masses \cite{NewALPSResult}. ALPS I used a single optical cavity placed before a light tight barrier to increase the circulating power on the axion production side of the magnet. The second iteration of the experiment, ALPS II, will improve the sensitivity further with the addition of an optical cavity after the barrier. The presence of this cavity will resonantly enhance the probability that axions/axion-like particles will reconvert to photons \cite{HOOGEVEEN19913, FUKUDA1996363, SikivieEnhanced, Mueller2009, Bahre2013}. ALPS II is currently being developed in two stages.  The first stage, ALPS IIa, will use two $10 \, \textrm{m}$ long resonant cavities without magnets \cite{Spector16}. The second stage, ALPS IIc, will use two 100 m long cavities with $5.3 \, \textrm{T}$ superconducting HERA dipole magnets. Longer cavity lengths increase the interaction time between the photons and the magnetic field. 

\begin{figure}[h]
\centering
\includegraphics[width=9cm]{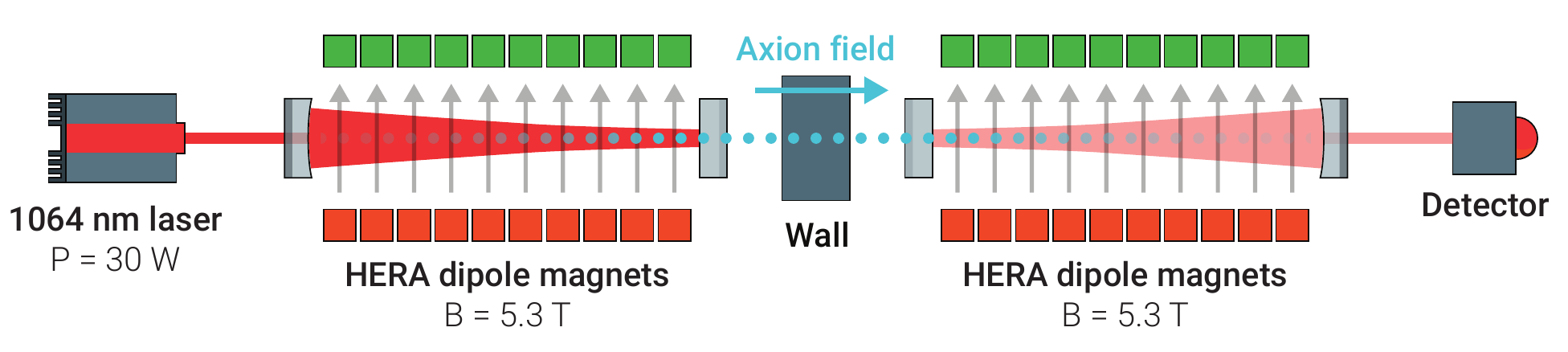}
\caption{Simplified model of the ALPS IIc experiment. Axions generated in the left-hand side cavity (the production cavity) traverse the wall and some turn back into detectable photons in the right-hand side cavity (the photon regeneration cavity).}
\label {fig:ALPSGeneral}
\end{figure} 

Figure~\ref{fig:ALPSGeneral} shows a simplified layout of the ALPS IIc experiment. Infrared laser light is injected into an optical cavity whose eigenmode is immersed in a $5.3 \, \textrm{T}$ magnetic field. The polarization of the injected light is set to be linear. The direction of the polarization is oriented either parallel or orthogonal to the direction of the external magnetic field in order to search separately for pseudo-scalar or scalar axion-like particles. Power buildup from this cavity causes a high circulating laser power, increasing the flux of axion-like particles through the wall. After these particles traverse the light-tight barrier they enter a second cavity, called the regeneration cavity, also subject to a $5.3 \, \textrm{T}$ magnetic field. The particles then have the same probability to reconvert back into photons having an energy identical to those in the production cavity.

\begin{figure}[h]
\centering
\includegraphics[width=9cm]{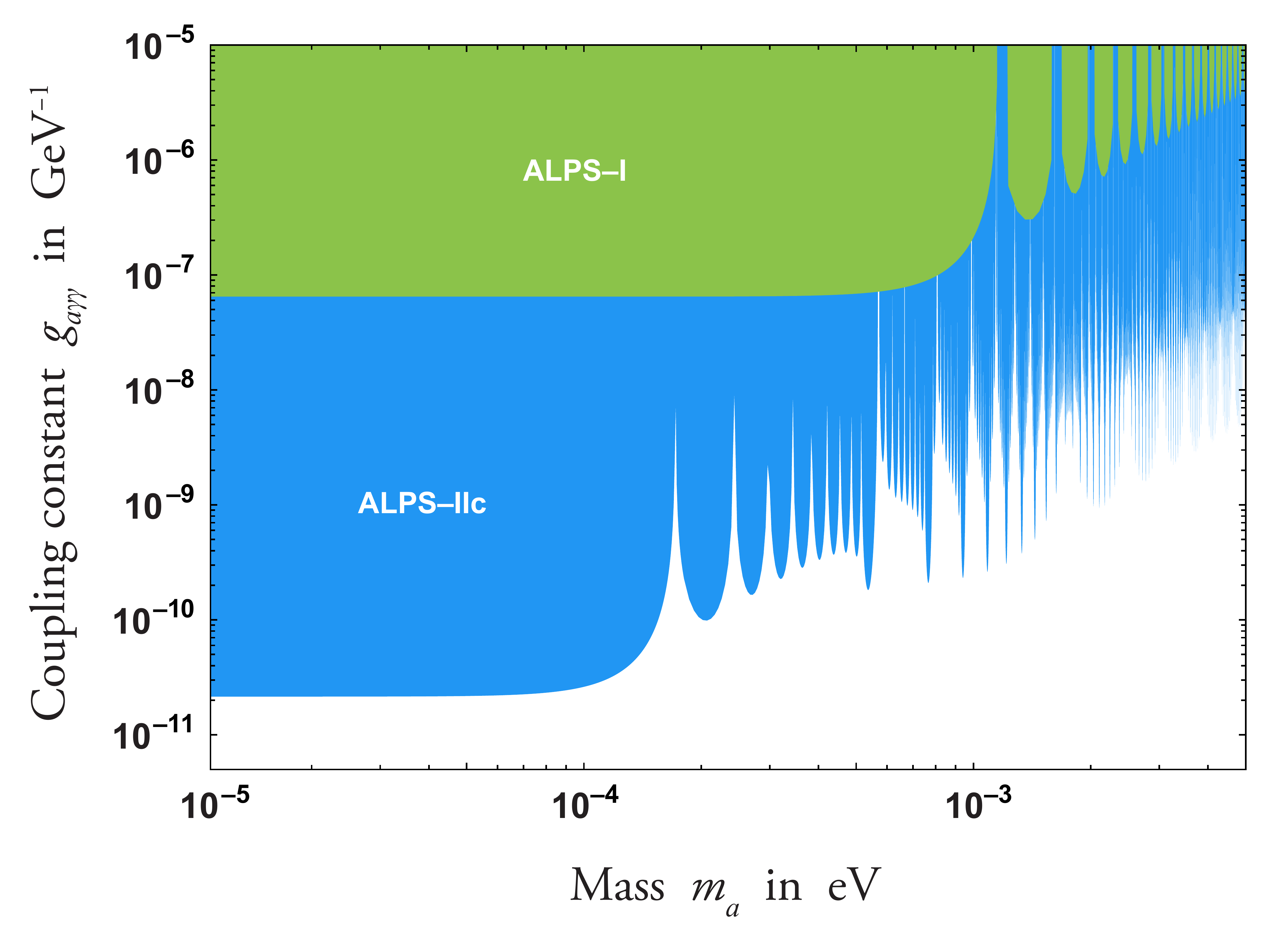}
\caption{Parameter space of the axion mass ($m_a$) and coupling ($g_{a \gamma \gamma}$) showing projected improvements in sensitivity from ALPS I (in vacuum) to ALPS IIc \cite{Bahre2013}.}
\label {fig:Improvements}
\end{figure}

The exclusion limits (95\% confidence level) measured by ALPS I for a 31 hour data run in vacuum is shown by the green region of Fig.~\ref{fig:Improvements}. Improvements in the optical design from ALPS I to ALPS IIc lead to a projected 2000-fold increase in sensitivity to the coupling parameter, $g_{a \gamma \gamma}$ . 
 

ALPS IIc will inject a 30 W laser field into the 100 m long production cavity (PC) which is immersed in a $5.3 \, \textrm{T}$ magnetic field. The circulating power inside the PC is expected to reach 150 kW. The 100 m long regeneration cavity (RC) on the other side of the wall will have a finesse of 120,000. The RC is also placed inside a similar 5.3 T magnetic field. Assuming a coupling strength of $g_{a \gamma \gamma} \sim 2 \times 10^{-11} \, \textrm{GeV}^{-1}$, the employed photon detector has to be able to measure fields with a photon rate as low as 2$\times 10^{-5}$ photons per second \cite{Mueller2009}. For 1064 nm light, this is equivalent to an average power on the order of $10^{-24} \, \textrm{W}$. ALPS II is exploring two technologies for detecting such weak signals. The first of these uses a transition edge sensor \cite{TES}. This technology utilizes a superconducting thin film operating near its critical temperature to absorb the regenerated photons, thereby changing its temperature and thus its resistance. An alternative approach, heterodyne interferometry, is the subject of this paper \cite{Mueller2009}.

 
\subsection{Heterodyne Detection Principles}
The principle of heterodyne interferometry requires interfering two laser fields at a non-zero difference frequency. Let one laser, $L_1,$ have frequency $f$, phase $\phi_1$, and average power $\bar{P}_{\textrm{1}}$ and a second laser, $L_2$, have frequency $f+ f_0$, phase $\phi_2$, and average power $\bar{P}_{\textrm{2}}$. Optically mixing these lasers at a photodetector yields the following expression:
\begin{multline}
\label{SimpleMixing}
\left | \sqrt{\bar{P}_{\textrm{1}}} e^{i  (2 \pi f t + \phi_1)} + \sqrt{\bar{P}_{\textrm{2}}} e^{i  [2 \pi (f+f_0) t + \phi_2]} \right |^2 = \\
 \bar{P}_{\textrm{1}} + \bar{P}_{\textrm{2}} + 2 \sqrt{\bar{P}_{\textrm{1}}\bar{P}_{\textrm{2}}} \cos{(2 \pi f_0 t + \Delta \phi)}~.
\end{multline}
Here, we have written the laser field amplitudes as proportional to the square root of the average power and have set $\Delta \phi = \phi_2 - \phi_1$. 
While the first two terms on the right side of the equation are the individual DC powers, the third term is a time varying signal, called a beat note, at the difference frequency, $f_0$. 

In our implementation of the heterodyne readout, the detector photocurrent, represented by Eq.~\ref{SimpleMixing}, is digitized satisfying the Nyquist criterion for sampling signals at $f_0$. The band-limited signal is then mixed to an intermediate frequency and written to file using a Field Programmable Gate Array (FPGA) card. Then, a second mixing stage in post-processing shifts the signal to DC, splitting it into two quadratures. Each resultant quadrature is continuously integrated over the measurement time. In order for the signal to accumulate, phase coherence between the two laser fields must be maintained during this entire process. The two quadratures are then combined to give a single quantity proportional to the product of the photon rate of each laser.

Implementation of a heterodyne detection scheme in ALPS II will involve injecting a second laser, phase coherent with the signal field and resonant in the regeneration cavity at a known offset frequency. The overlapped beams are transmitted out of the cavity and are incident onto the heterodyne detector.

In this report we present results from a test setup which validates the approach and will guide its implementation in ALPS IIc. To begin, in Section~\ref{Section2} we mathematically demonstrate how a coherent signal is extracted from the input. In Section~\ref{Section3}, we then discuss the optical design created to test this technique and the digital design which forms the core of heterodyne detection. Finally, in Section~\ref{Section4} we present results on device sensitivity and coherent signal measurements.

\section{Mathematical Expectations} \label{Section2}
\subsection{Signal Behavior}
In our standalone experiment, two lasers are interfered and incident onto a photodetector. Laser 1 acts as our local oscillator (LO) with average power $\bar{P}_{\textrm{LO}}$ while Laser 2 provides the signal field we wish to measure with average power $\bar{P}_{\textrm{signal}}$. The difference frequency is set such that the generated beat note has frequency $f_{\textrm{sig}}$. Once the combined beam is incident onto a photodetector with gain $G$ in ${\textrm{V}}/{\textrm{W}}$, it is digitized into discrete samples, $x[n]$, where $n$ is the individual sample number, at sampling frequency $f_s$. Sampling is done using an analog-to-digital converter (ADC) with a 1 V reference voltage. In the absence of noise, the AC component becomes
\begin{equation}
\label{SignalForm}
x_{\textrm{sig}}[n] = 2 G \sqrt{\bar{P}_{\textrm{LO}} \bar{P}_{\textrm{signal}}} \cos{(2\pi \frac{f_{\textrm{sig}}}{f_s} n + \phi)}~,
\end{equation}
where $\phi$ is an unknown but constant phase. 

In order to recover amplitude information, the digitized beat note signal is separately mixed with a cosine/sine wave at frequency $f_d = f_{\textrm{sig}}$ in a process known as I/Q demodulation:
\begin{equation}
\label{SignalOnlyIQ}
\begin{split}
&I[n] = x_{\textrm{sig}}[n]  \times \cos{(2\pi \frac{f_d}{f_s} n)} \\
&Q[n] = x_{\textrm{sig}}[n] \times \sin{(2 \pi \frac{f_d}{f_s} n)}~. \\
\end{split}
\end{equation}
Each quadrature is individually summed from $n = 1$ to $N$ samples. The squared sums are added together and normalization is done through division by $N^2$. This entire quantity is given by the following expression,
\begin{equation}
\label{RootT}
Z(N) = \frac{(\sum_{n=1}^N I[n])^2 + (\sum_{n=1}^N Q[n])^2}{N^2}~.
\end{equation}
The numerator is in fact the square of the magnitude of the discrete Fourier transform (DFT) of the digitized input\footnote{It must be noted that this is only exactly true in the case that $\frac{f_d}{f_s} = \frac{k}{N}$ for some integer $k$. If this requirement is not met then the windowing process results in spectral leakage and $Z(N)$ becomes an estimate of the DFT. However, in the large $N$ limit this leakage becomes negligible.} evaluated at ${f_d}/{f_s}$: 
\begin{equation}
\label{ZDFT}
Z(N) = \frac{|X\left [\frac{f_d}{f_s}\right ]|^2}{N^2}~,
\end{equation}
where
\begin{equation}
\label{DFTdef}
X\left [\frac{f_d}{f_s}\right ] = \sum_{n=1}^{N}x[n]e^{-i2\pi \frac{f_d}{f_s} (n-1)}~.
\end{equation}
Setting $f_d = f_{\textrm{sig}}$ and solving for $Z(N)$ with an input given by Eq.~\ref{SignalForm} yields,
\begin{equation}
\label{SignalYield}
Z_{\textrm{sig}}(N) = G^2 \bar{P}_\textrm{{LO}} \bar{P}_{\textrm{signal}}~.
\end{equation}

Demodulating at the beat note signal frequency causes $Z(N)$ to be proportional to $\bar{P}_{\textrm{signal}}$ and constant with integration time, $\tau$. The power in the local oscillator amplifies the beat note amplitude and will be set to overcome all technical noise sources.

\subsection{Noise Behavior}
We wish to set the signal field to compare with the projected sensitivity of ALPS IIc on the order of a few photons per week. Therefore, we must consider the influence of important noise sources such as laser relative intensity noise and optical shot noise. In order to understand the influence of such noise, let us determine $Z(N)$ in the absence of an RF signal ($\bar{P}_{\textrm{signal}} = 0$) but in the presence of noise. 

Consider the input $x[n]$ to be a random stationary process. The quantity $Z_{\textrm{noise}}(N)$ can be written in terms of the single-sided analog power spectral density (PSD) evaluated at the demodulation frequency, $f_d$. To do so, we first convert the analog PSD in ${\textrm{V}^2}/({\textrm{cycles per second}})$ to the digitized power spectral density (DPSD) in ${\textrm{V}^2}/({\textrm{cycles per sample}}) $ using the sampling frequency $f_s$ \cite{stoica2005spectral}. 
\begin{equation}
\label{PSDtoSampled}
\textrm{DPSD}\left (\frac{f_d}{f_s}\right )  = f_s \ \textrm{PSD}(f_d) 
\end{equation}
The DPSD is related to the expectation, $\mathcal{E}$, of the DFT of $x[n]$ \cite{stoica2005spectral}:
\begin{equation}
\label{PSDtoDFT}
\textrm{DPSD}\left (\frac{f_d}{f_s}\right ) = \lim_{N \rightarrow \infty} \mathcal{E}\left [ \frac{|X\left [\frac{f_d}{f_s}\right ]|^2}{N} \right ]~.
\end{equation}
Using Eq.~\ref{ZDFT} we can solve for $Z(N)$.
\begin{equation}
\label{ZNPSD}
\lim_{N \rightarrow \infty} \mathcal{E}\left [ Z_{\textrm{noise}}(N) \right ] = \frac{\textrm{PSD}(f_d)}{\tau}~,
\end{equation}
where we use the substitution $N = \tau f_s$. It is important to note that this only depends upon the PSD evaluated at $f_d$ and not across the entire spectrum.

Although Eq.~\ref{ZNPSD} exactly relates the expectation value of $Z_{\textrm{noise}}(N)$ to the analog PSD, we are interested in the result of a single trial. For such an individual trial, $Z_{\textrm{noise}}(N)$ provides only an estimate of the analog PSD. Because the noise is assumed to be stationary, the PSD is by definition constant with time. The behavior of $Z_{\textrm{noise}}(N)$ for a single trial therefore tends to fall off as 1/$\tau$. However, for a set integration time the outcome of multiple trials of $Z_{\textrm{noise}}(N)$ will have some non-zero variance \cite{stoica2005spectral, peligrad2010}. 
\begin{equation}
\lim_{N \rightarrow \infty} \sigma_{Z}^2 = \left ( \frac{\textrm{PSD}(f_d) }{\tau} \right )^2
\end{equation}

A confidence threshold for a single run must therefore be determined in order to distinguish between coherent detection of a signal and the random nature of this noise. From this point forward we assume $N$ to be sufficiently large such that Eq.~\ref{ZNPSD} and its derivatives provide good approximations to real world applications.  

\subsection{Detection Threshold}
To simplify this calculation let us assume that the input is appropriately band-pass filtered around $f_d$ and downsampled such that the resulting frequency spectrum is locally flat. It has been shown that in the large $N$ limit $X \left[ {f_d}/{f_s} \right ]$ is a Gaussian random variable, independent of the other $X \left [{f}/{f_s} \right ]$ due to the central limit theorem \cite{peligrad2010}. $Z_{\textrm{noise}}(N)$ therefore follows an exponential distribution. Using the cumulative distribution function \cite{papoulis2002probability}, the probability, $\mathcal{P}$, of measuring a final value of $Z_{\textrm{noise}}(N)$ between 0 and an upper limit $u$ for a given $\tau$ is 
\begin{equation}
\label{cdf}
\mathcal{P}(u) = 1- e^{-u/\sigma_{Z}}~.
\end{equation}
From the inverse of Eq.~\ref{cdf}, we can define a probability range for individual outcomes of $Z_{\textrm{noise}}(N)$ to fall between 0 and an upper limit for any given probability $\mathcal{P}$. 
For the 5-sigma limit ($\mathcal{P}_{5s} = 0.9999994$) this is
\begin{equation}
\label{5sigmaPSDcalc}
u(\mathcal{P}_{5s})[Z_{\textrm{noise}}(N)] = -\textrm{ln}(6 \times 10^{-7}) \frac{\textrm{PSD}(f_d)}{\tau}~.
\end{equation}
Consequently, when $Z(N)$ has a value above this limit for a predefined number of samples $N$, we can claim with 99.99994\% confidence that a coherent signal is present. 

\begin{figure}[t]
\centering
\includegraphics[width=9cm]{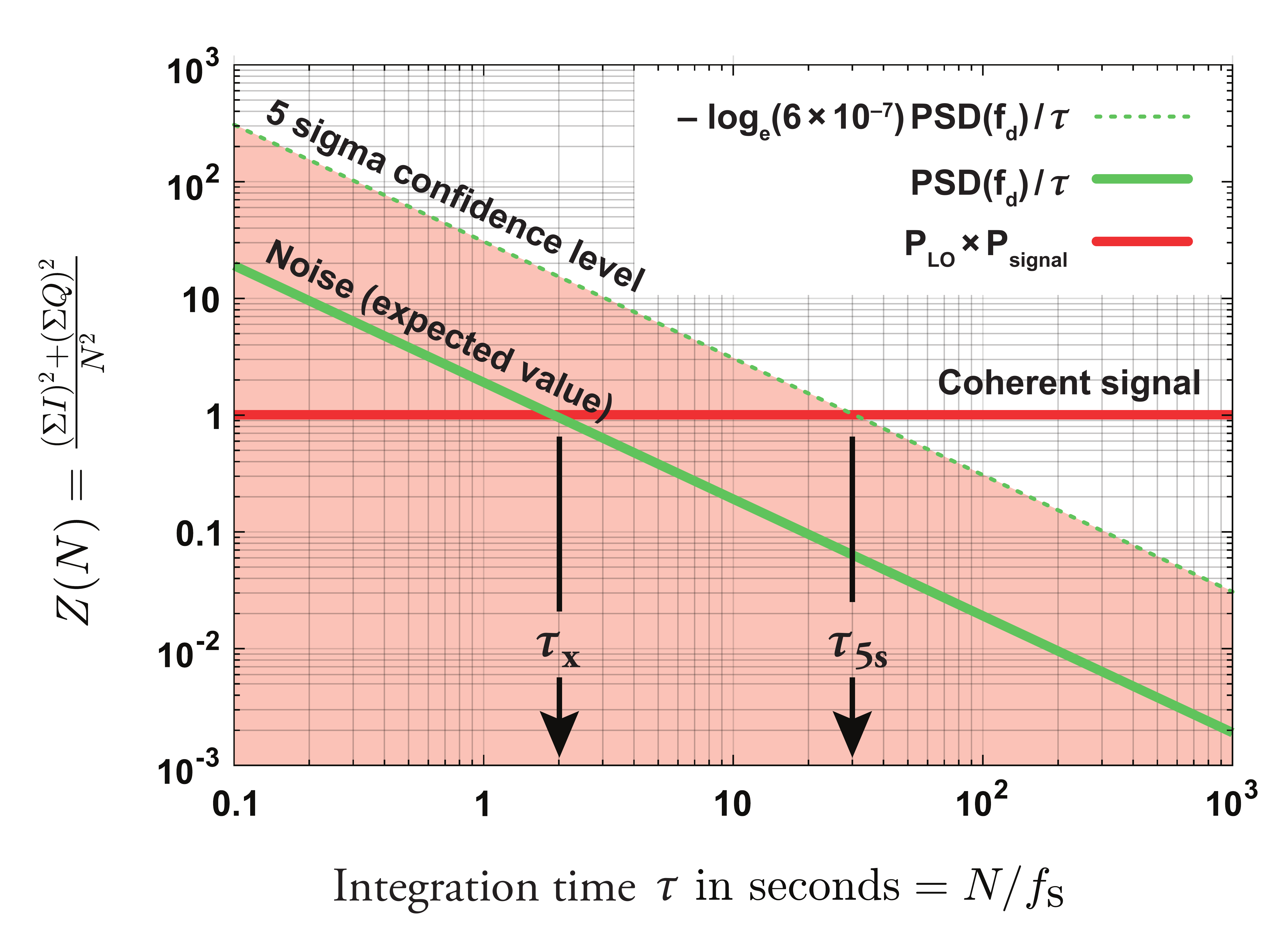}
\caption{Expected behavior of noise, signal, and the 5-sigma limit when plotting $Z(N)$ vs. integration time $\tau$. Noise and the 5-sigma limit both go as 1/$\tau$ whereas the signal stays flat with time. Because $Z(N)$ is proportional to the power in the signal field we can scale the y-axis accordingly using the gain factors within our system in order to obtain a meaningful photon rate of the weak field. Noise-level-dependent integration times $\tau_{\text{x}}$ (where the signal crosses the expected value of noise) and $\tau_{\textrm{5s}}$ (where a detection can be claimed with 5-sigma confidence) are highlighted.}
\label {fig:ZachSquared}
\end{figure} 

The expected behaviors of $Z(N)$ and the 5-sigma limit are plotted vs integration time $\tau$ in Fig.~\ref{fig:ZachSquared}. When a beat note signal is present at frequency $f_{\textrm{sig}} = f_d$  (Eq.~\ref{SignalYield}), the expectation value, shown in red, is constant with integration time and scales linearly with the power of the signal field, $\bar{P}_{\textrm{signal}}$. This power can be expressed in terms of photons per second, our quantity of interest. 

Following Eq.~\ref{ZNPSD}, the expectation value of $Z_{\textrm{noise}}(N)$ (signal absent), shown as the solid green line, goes as 1/$\tau$. Similarly the 5-sigma limit falls off as 1/$\tau$ according to Eq.~\ref{5sigmaPSDcalc}.

\subsection{Fundamental limits}
From now on, we will scale $Z_{\textrm{signal}}(N)$ to photons per second in the signal field, $\bar{P}_{\textrm{signal}}/h\nu$. A scaling factor of $1/(G^2 h\nu \bar{P}_{\textrm{LO}})$ is applied to Eq.~\ref{SignalYield} such that
\begin{equation}
\label{4264634}
 \frac{Z_{\textrm{signal}}(N)}{G^2 h\nu \bar{P}_{\textrm{LO}}} = \frac{\bar{P}_{\textrm{signal}}}{h\nu}~.
\end{equation}
where $h$ is the Planck constant and $\nu$ is the laser frequency, so that $h\nu$ is the photon energy. Scaling the noise (Eq.~\ref{ZNPSD}) and 5-sigma limit (Eq.~\ref{5sigmaPSDcalc}) by the same factor yields
\begin{equation}
\label{ZNPSDPho}
\frac{\mathcal{E}\left [ Z_{\textrm{noise}}(N) \right ]}{G^2 h\nu \bar{P}_{\textrm{LO}}} = \frac{\textrm{PSD}(f_d)}{G^2 h\nu \bar{P}_{\textrm{LO}}\times \tau }~,
\end{equation}
and
\begin{equation}
\label{5sPhotons}
\frac{u(P_{\textrm{5s}})}{G^2 h\nu \bar{P}_{\textrm{LO}}} = \frac{ -\textrm{ln}(6 \times 10^{-7}) \ \textrm{PSD}(f_d)}{G^2 h\nu \bar{P}_{\textrm{LO}} \times \tau }~.
\end{equation}

The fundamental noise source in our optical heterodyne detection setup as well as in ALPS IIc is shot noise (sn). This type of noise is well characterized and follows Poisson statistics \cite{fox}. Experimentally we ensure that our system is shot-noise limited at the demodulation frequency. We may then use the known PSD for shot noise in ${\textrm{A}^2}/{\textrm{Hz}}$ \cite{shotnoiseEq},
\begin{equation}
\label{PSDsn}
\textrm{PSD}_{\textrm{sn}}  = 2 q {I}_{\textrm{DC}}~,
\end{equation} 
where $q$ is the electron charge. The DC photocurrent, ${I}_{\textrm{DC}}$, is related to the total input average optical power. With $\bar{P}_{\textrm{LO}} \gg \bar{P}_{\textrm{signal}}$ the photocurrent becomes,
\begin{equation}
\label{IDC}
{I}_{\textrm{DC}} = \eta \frac{q}{h \nu} \bar{P}_{\textrm{LO}}~,
\end{equation}
where $\eta$ is the quantum efficiency of the photodetector. 
Finally, we use the photodetector gain $G$ in order to convert this to ${\textrm{V}^2}/{\textrm{Hz}}$.
\begin{equation}
\label{PSDFinal}
\textrm{PSD}_{\textrm{sn}} = 2 G^2 {h \nu } \bar{P}_{\textrm{LO}} \frac{1}{{\eta}}
\end{equation}

Substituting this quantity into Eq.~\ref{ZNPSDPho} yields the expected behavior when shot noise is the dominant source at the demodulation frequency. 
\begin{equation}
\label{ShotNoisePhotonLimitSingle}
\frac{\mathcal{E}\left [ Z_{\textrm{sn}}(N) \right ]}{G^2 h\nu \bar{P}_{\textrm{LO}}} = \frac{2}{\eta \tau}
\end{equation}

Because the left-hand side of this equation is equal to the photon rate of the signal field if a signal is present, using Eq.~\ref{4264634} we can predict the time at which a signal will cross the expected value of this fundamental noise limit,
\begin{equation}
\label{TauXSN}
\tau_{\textrm{x,sn}} = 2\frac{h\nu}{\eta \bar{P}_{\textrm{signal}}}~.
\end{equation}
Similarly from Eq.~\ref{5sPhotons}, we find that the time required for the signal to cross the 5-sigma detection threshold is
\begin{equation}
\label{TimeCross5sPhotons}
\tau_{\textrm{5s,sn}} = -2 \ \textrm{ln}(6 \times 10^{-7})  \frac{h\nu}{\eta \bar{P}_{\textrm{signal}}} \approx 29  \frac{h\nu}{\eta \bar{P}_{\textrm{signal}}}~,
\end{equation}
in the case of a shot-noise limited input. 

In conclusion, for a weak field with a power equivalent to 1 photon per second it takes 2 seconds for the expected value of shot noise to decrease to the signal level with $\eta = 1$. However, it takes $\sim 29$ seconds in order to claim a detection of a signal with 5-sigma confidence.
For arbitrary noise input, both integration times, as depicted in Fig.~\ref{fig:ZachSquared}, can be generalized to
\begin{equation}
\label{TauX}
\tau_{\textrm{x}} = \frac{\textrm{PSD}(f_d)}{ G^2} \times \frac{1}{\bar{P}_{\textrm{LO}} \bar{P}_{\textrm{signal}}}\quad~,
\end{equation}
and
\begin{equation}
\label{Tau5s}
\tau_{\textrm{5s}} = \frac{\textrm{PSD}(f_d)}{ G^2}  \times \frac{- \textrm{ln}\left(6 \times 10^{-7}\right)}{\bar{P}_{\textrm{LO}} \bar{P}_{\textrm{signal}}}~,
\end{equation}
if the noise is locally flat around $f_d$.
The factor between $\tau_{\textrm{5s}}$ and $\tau_{\textrm{x}}$
\begin{equation}
\label{TauRatio}
\frac{\tau_{\textrm{5s}}}{\tau_{\textrm{x}}} =  -\ \textrm{ln}\left(6 \times 10^{-7}\right) \approx 14~,
\end{equation}
does not depend on the PSD, the average power of either laser, or the sampling frequency $f_s$. 

Additionally, Eq.~\ref{TauX} shows the importance of a higher power $\bar{P}_{\textrm{LO}}$ when the system is not dominated by shot noise. The larger the LO power, the less time required for the signal to cross the expected noise limit, thus improving the SNR. However, once $\bar{P}_{\textrm{LO}}$ is large enough such that the system is shot-noise limited, $\tau_{\textrm{x}}$ and, consequently, the SNR no longer depend on the LO power.

\section{Experimental Setup} \label{Section3}
\subsection{Optical Design}
\begin{figure}[h]
\centering
\includegraphics[width=9cm, page = 1]{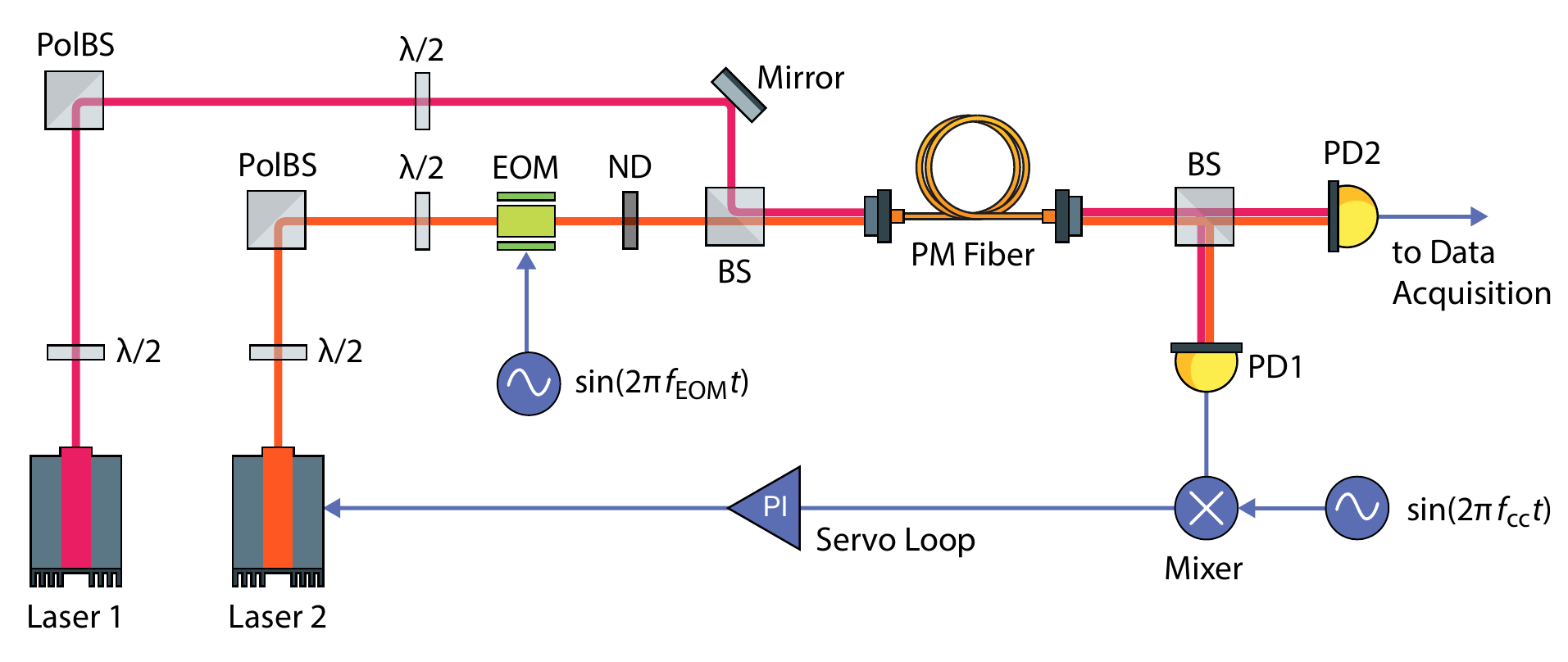}
\caption{Optical layout of the heterodyne interferometer used for single photon detection.$\ \lambda$/2 denotes a half-wave plate, PolBS refers to a polarizing beam splitter, BS denotes a 50/50 power beam splitter, ND refers to the neutral density filters, EOM is the electro-optic modulator, PM fiber is the polarization maintaining optical fiber, and PD is a photodetector.}

\label {fig:OpticalSetup}
\end{figure}

To demonstrate this concept experimentally, we assembled the optical setup shown in Fig~\ref{fig:OpticalSetup}. This apparatus allows us to measure the resultant beat note generated from interfering a weak signal field with our LO. 
There are two 1064 nm lasers used, $L_1$ and $L_2$.$\ L_1$ is our LO and $L_2$ provides the field used for weak signal generation. A half-wave plate and polarizing beam splitter (PolBS) pair is placed at the start of each beam path for power control purposes. This combination also causes the outgoing light to be linearly polarized. Laser 2 passes through an electro-optic modulator (EOM) which generates sidebands to be used as the weak signal. This will be discussed in more detail later in this section. Laser 2 then passes through two neutral density (ND) filters with a combined attenuation factor of $\sim 2 \times 10^5$ in order to further reduce the weak-field signal to an appropriate level. 


The two fields are interfered at a 50/50 power beam splitter (BS) and the combined beam is sent into a single-mode polarization-maintaining fiber. By sending both beams into the same single-mode fiber we ensure complete overlap of the spatial eigenmodes at the output coupler. After the fiber, the combined beam passes through another 50/50 power BS. Each path is then focused individually onto separate photodetectors. PD1 is used to lock the two lasers to the constant difference frequency. This is done via feedback to the laser controller for $L_2$ using a phase lock loop (PLL) setup. PD2 is a homemade photodetector used for our signal measurements. For a large enough local oscillator power the shot noise level exceeds the noise equivalent power (NEP) of the photodetector and PD2 produces shot-noise limited signals. We set the average local oscillator power to 5 mW and observe a shot noise to NEP ratio of 6 at the measurement frequency. 

Overlapping the two lasers generates a beat note between $L_1$ and $L_2$, called the carrier-carrier (CC) beat note at frequency $f_{\textrm{CC}}$. The error signal for the PLL feedback comes from mixing the carrier-carrier beat note with a numerically controlled oscillator (NCO), also at frequency $f_{\textrm{CC}}$, synchronized to a master clock. This feedback is controlled by the FPGA card and keeps the CC beat note stable at frequency $f_{\textrm{CC}}$. 


\subsection{Digital Design}
\begin{figure}[h]
\centering
\includegraphics[width=9cm, page = 2]{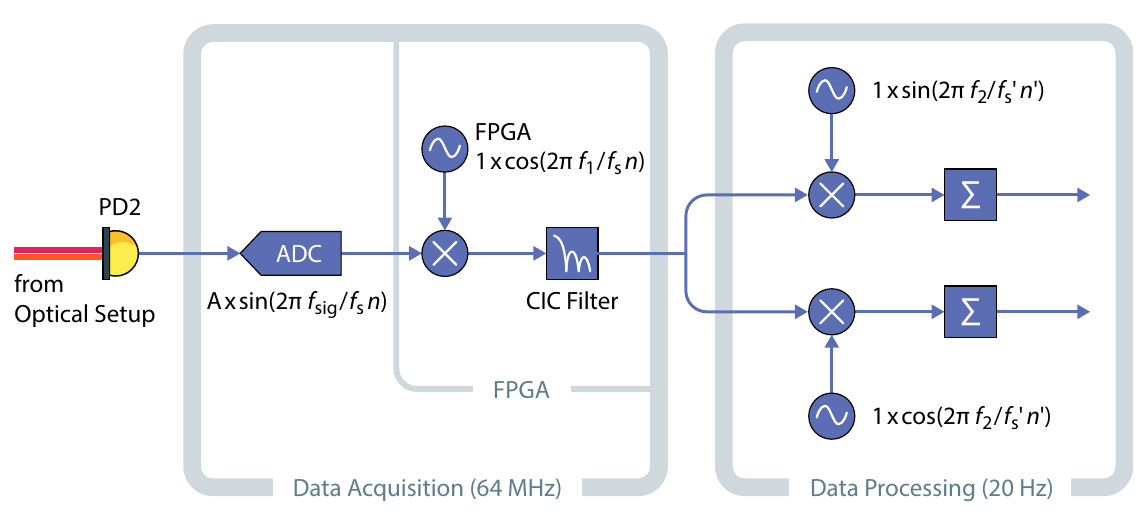}
\caption{Digital layout of detection scheme describing the digital processing techniques involved. The photodetector signal is digitized via an analog-to-digital converter at a rate of 64 MHz after which it is mixed with a sine wave, produced by a numerically controlled oscillator, at frequency $f_1$. A cascaded integrated comb filter is used to remove the higher frequency components due to mixing and downsample the data to 20 Hz, where is it written to file. $f_s'$ and $n'$ are used to reference the lower sampling rate. I/Q demodulation is done onboard a desktop computer, and the quadratures are individually summed and $Z(N)$ is computed.}
\label {fig:ElectricalSetup}
\end{figure}

The electrical signal from PD2 is digitized via an ADC on-board an FPGA card at a rate of $f_s =64$ MHz. A simplified digital layout following the path of the photodetector signal is detailed in Fig.~\ref{fig:ElectricalSetup}.

The signal at frequency $f_{\textrm{sig}}$  is mixed down to an intermediate frequency, $f_{\delta}$, on the order of a few Hz. This is done via multiplication with a sinusoid from an NCO at frequency $f_1 = f_{\textrm{sig}} - f_{\delta}$ generated with a look-up table on the FPGA card.

While it is possible to demodulate directly down to DC during the first demodulation stage simply by setting the NCO frequency to $f_1 = f_{\textrm{sig}}$, we observed spurious DC signals generated within the FPGA card when tested with this configuration. The strength of these signals are orders of magnitude larger than the beat notes of interest thus preventing any useful measurements. This issue is solved by mixing the beat note signal down to the intermediate frequency, writing the data to file, and performing a second demodulation stage on a desktop PC. This double demodulation shifts the unwanted spurious signal to a non-zero frequency where it integrates away. With this configuration, the beat note can be accurately measured.

A cascaded integrated comb (CIC) filter \cite{Hogenauer1981}, removes the higher frequency components resulting from the mixing process. The CIC filter also downsamples the data to a rate of $f_s' \approx 20$ Hz at which they are written to file.

The signal at $f_{\delta}$ is decomposed into its in-phase (I) and quadrature (Q) components via separate mixing with a cosine and sine NCO at $f_2 = f_{\delta}$, respectively. Considering the same signal input from Eq.~\ref{SignalForm}, this process referenced to the higher sampling rate, $f_s$, is equivalent to: 
\begin{equation}
\label{IDemodSig}
\begin{split}
&I_2[n] = x_{\textrm{sig}}[n]  \sin{(2\pi \frac{ f_1}{f_s}n)} \times \cos{( 2\pi \frac{f_2 }{f_s}n)} \\
&Q_2[n] = x_{\textrm{sig}}[n]  \sin{(2\pi  \frac{f_1 }{f_s}n )}  \times \sin{(2\pi  \frac{f_2 }{f_s}n)}~. \\
\end{split}
\end{equation}
The DFT of the recorded data at the lower sampling rate, $|X \left [{f_2}/{f_s'} \right ]|^2$, is then computed. The expectation values of $Z(N)$ from Section~\ref{Section2} must be rewritten to include this second demodulation stage. We denote these equations with a subscript 2. The total number of samples written to file is $N' = \tau f_s'$.  

With a signal present at the demodulation frequency in the absence of noise we find
\begin{equation}
\label{SBPower2}
Z_{2,\textrm{sig}}(N') = \frac{G^2}{4} \bar{P}_{\textrm{LO}} \bar{P}_{\textrm{signal}}~.
\end{equation}
The photon rate in the signal field is
\begin{equation}
\label{Z2Sig2}
\frac{4 \ Z_{2,\textrm{sig}}(N')}{G^2 h\nu \bar{P}_{\textrm{LO}}} = \frac{\bar{P}_{\textrm{signal}}}{h\nu}~.
\end{equation}
Using this new scaling factor of 4/$(G^2 h \nu \bar{P}_{\textrm{LO}})$, we obtain a quantity equal to the photon rate of the signal field after two demodulation stages. 

In the case where there is only noise at the input, the PSD when the data are recorded ($\textrm{DPSD}^\prime$) must be related to the original DPSD right after the ADC. Multiplication by a sine wave reduces the DPSD by a factor of 2. The decimation stage reduces the level of the DPSD by a factor of ${f_s'}/{f_s}$. For $|f_2| \leq {f_s'}/{2}$, 
\begin{equation}
\textrm{DPSD}^\prime \left (\frac{f_2}{f_s'}\right ) = \frac{1}{2} \  \frac{f_s'}{f_s} \  \textrm{DPSD} \left (\frac{f_d}{f_s}\right ) = \frac{f_s'}{2} \  \textrm{PSD}(f_d)~.
\end{equation}
This quantity is related to the DFT by
\begin{equation}
\textrm{DPSD}^\prime \left (\frac{f_2}{f_s'}\right ) = \mathcal{E} \left \{ \frac{|X \left [\frac{f_2}{f_s'} \right ]|^2} {N'} \right \} = \mathcal{E} \{ {Z_{2 }(N')} \times N' \}~.
\end{equation}
Solving for $\mathcal{E}[Z_{2}(N')]$ in terms of the analog PSD evaluated at $f_d = f_1 + f_2$ gives 
\begin{equation}
\label{Noise2}
\mathcal{E}[Z_{2,\textrm{noise}}(N')] =  \frac{\textrm{PSD}(f_d)}{{2 \tau}}~,
\end{equation}
where we use the substitution $N' = \tau f_s'$. In order to compare the expectation value of noise to that of the signal, we must apply the new scaling factor of 4/$(G^2 h \nu \bar{P}{\textrm{LO}})$. Doing so, we arrive at
\begin{equation}
\label{NoisePho2}
\frac{4 \ \mathcal{E}[Z_{2,\textrm{noise}}(N')]}{G^2 h\nu \bar{P}_{\textrm{LO}}} = \frac{2 \  \textrm{PSD}(f_d)}{G^2 h\nu \bar{P}_{\textrm{LO}} \times \tau}~.
\end{equation}
For the shot-noise-limited case with $\textrm{PSD}_{\textrm{sn}}$ given by Eq.~\ref{PSDFinal}, this calculation yields
\begin{equation}
\label{NoisePho2SN}
\frac{4 \ \mathcal{E}[Z_{2,\textrm{sn}}(N')]}{G^2 h\nu \bar{P}_{\textrm{LO}}} = \frac{4}{\eta \tau}~.
\end{equation}
Comparing to Eq.~\ref{ShotNoisePhotonLimitSingle}, the introduction of a second demodulation stage causes the sensitivity to decrease by a factor of 2. This decrease, in turn, also causes the 5-sigma limit to increase by a factor of 2. Therefore, using two demodulation stages requires twice as long an integration time (when compared to a single stage setup) to detect confidently a signal.\footnote{In principle, it is possible to regain the earlier sensitivity while still using two demodulation stages. This is done by taking both I and Q out of the FPGA. Then a second I/Q demodulation is performed on each output channel. This results in four terms II', IQ', QI', and QQ' where the prime indicates the second demodulation stage. Using a specific combination of these terms yields the same set of equations described in Section~\ref{Section2} \cite{RDS}. This concept is currently being tested and has not yet been implemented.}

Signal and noise add linearly in the PSD:
\begin{equation}
\label{TotalZN}
\frac{4 \ \mathcal{E}[Z_{2,\textrm{total}}(N')]}{G^2 h\nu \bar{P}_{\textrm{LO}}} = \frac{\bar{P}_{\textrm{signal}}}{h\nu} + \frac{4}{\eta \tau}~.
\end{equation}
For short integration times and a low photon rate,  ${4}/{\eta \tau}$ is the dominating term. After long enough integration, ${\bar{P}_{\textrm{signal}}}/{h\nu}$ becomes dominant causing the curve to remain constant with time. 

These equations now reflect the result of adding a second demodulation stage, however, one final experimental consideration must be taken into account. Simply lowering the power of Laser 2 to sub-photon per second levels reduces the CC beat note below the point at which the PLL becomes unstable. Experimentally, a stable lock can be maintained with $\bar{P}_{\textrm{LO}} = 5$ mW and $\bar{P}_{\textrm{L2}} = 60$ pW measured at PD1. This leads to a minimum CC beat note amplitude on the order of 1 $\mu$W. Increasing $\bar{P}_{\textrm{LO}}$ any further pushes the photodetector past the level at which it begins to saturate. Therefore, the minimum photon rate of Laser 2 at PD2, such that the PLL remains stable, is $3 \times 10^{8}$ photons per second. In order to generate signals with field strengths below this value, while still maintaining a stable PLL between the two lasers, we make use of phase modulation from an EOM.


\subsection{EOM Sideband Generation}
As mentioned earlier, the EOM shown in Fig.~\ref{fig:OpticalSetup} was used to generate sidebands on Laser 2. The EOM is driven at frequency $f_{\textrm{EOM}}$ using a sine wave from a function generator that is synchronized to a maser clock. This voltage modulates the phase of the beam as it passes through the EOM. Phase modulation generates sidebands both above and below the laser frequency. These sidebands occur at k integer multiples of the drive frequency, $f_{\textrm{EOM}}$. The amount of light power in the k$^{\textrm{th}}$ order sideband is \cite{saleh2013fundamentals}
\begin{equation}
\label{SBPower}
\bar{P}_{\textrm{SB,k}} = J_\textrm{k}(m)^2 \bar{P}_{\textrm{L2}}~,
\end{equation}
where $J_\textrm{k}(m)$ is the $\textrm{k}^{\textrm{th}}$ order Bessel function and $m$ is the modulation depth, dependent on the drive amplitude of the modulation. All of these sidebands beat with the LO to produce AC signals with peak amplitudes given by the following,
\begin{equation}
\label{SBBeat}
A_\textrm{k} = 2\sqrt{\bar{P}_{\textrm{SB,k}} \  \bar{P}_{\textrm{LO}}}~.
\end{equation}

The two ND filters directly after the EOM attenuate the power of Laser 2 and all of the subsequent sidebands by a factor of $\sim 2 \times 10^5$. The addition of these ND filters is necessary to reduce the sideband power of interest to the desired level. 

The power in the k$^{\textrm{th}}$ sideband, $\bar{P}_{\textrm{SB,k}}$, can be fine tuned by adjusting the drive amplitude to the EOM, thus changing the modulation depth, $m$. To set the modulation depth to a specific value, the two ND filters are removed such that both the CC and sideband beat notes are visible on a spectrum analyzer. The ratio between the two beat note amplitudes is adjusted in order to obtain the desired modulation depth. The ND filters are then placed back into the beam path. 

Using this configuration, the average power of Laser 2 is set to maintain a stable PLL. Higher order sidebands fall off in power to levels comparable to the projected sensitivity of ALPS IIc. The interference between these sidebands and the LO form beat notes at known, fixed frequencies. 


\section{Results} \label{Section4}
\subsection{Noise Behavior and Device Sensitivity}
\begin{figure}[h]
\centering
\includegraphics[width=9cm]{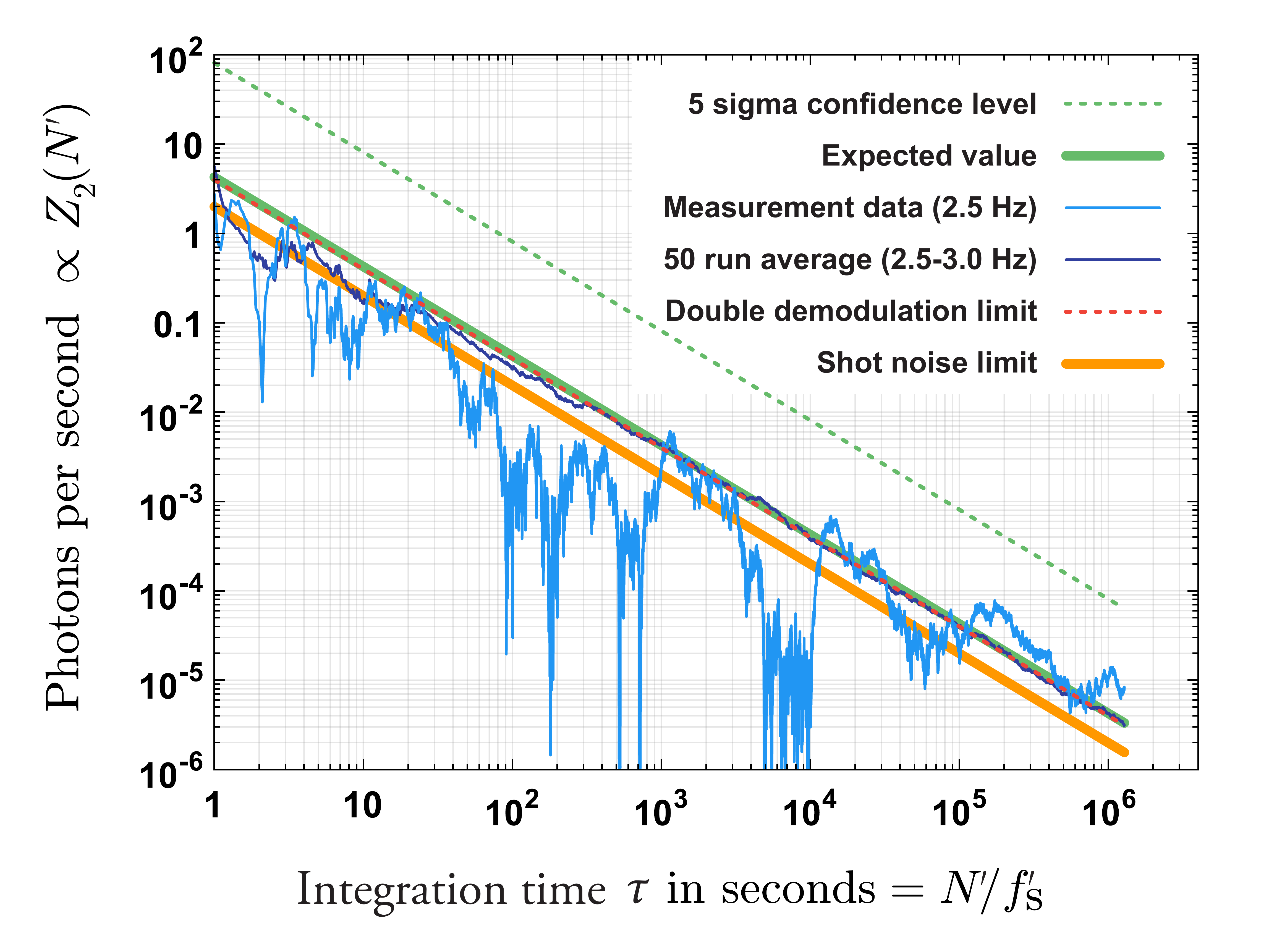}
\caption{Shot-noise limited measurement with no signal field present. After the second demodulation at $f_2 =$ 2.5 Hz, $Z_2(N')$ is computed and the resultant is scaled to an equivalent photon rate, shown in light blue. $Z_2(N')$ is also computed for 50 separate demodulation frequencies near 2.5 Hz. These data are then averaged to produce the dark blue curve. This average follows the expected value line, shown in solid green, based on the PSD of the noise.  The dashed green line shows the 5-sigma limit that the measurement curve would cross if a signal was present. The fundamental shot-noise limit (if only one demodulation stage was required) is drawn as the solid orange line for comparison. The second demodulation stage increases the shot-noise limit by a factor of 2 (dashed red). Because the expected value of the measurement sits on top of this theoretical limit we show that shot noise is the dominant noise source in our setup.}
\label {fig:2WeekNoise}
\end{figure}

We first performed a measurement with no signal field present to study the behavior of the noise in our system. Only the LO beam with power $\bar{P}_{\textrm{LO}} = 5$ mW is incident onto PD2. The photodetector is shot-noise limited at this level of incident light power. The photodetector has gain $G = 1.44 \times 10^{3} \ {\textrm{V}}/{\textrm{W}}$ and quantum efficiency $\eta = 0.7$. After both demodulation stages, $Z_2(N')$ is computed and the result is scaled to an equivalent photon rate using the factor stated in Eq.~\ref{NoisePho2SN}. The result of this 19-day measurement, plotted against integration time $\tau$, is shown in Fig.~\ref{fig:2WeekNoise}. 

$Z_2(N')$ was computed 50 additional times using different second demodulation frequencies near 2.5 Hz. The results are then scaled to the photon rate and averaged. This average is identical to the curve representing the expectation values for different integration times. Both have essentially the same amplitude and fall off as 1/$\tau$ as expected. The data stream demodulated at $f_2 = 2.5$ Hz shows one representation of a shot noise dominated signal over the integration time.  In addition, the 5-sigma threshold is plotted.

The solid orange line shows the expected fundamental shot-noise limit for the given optical power if only one demodulation stage was used. As our measurement requires a second demodulation stage, the amount of shot noise returned by the measurement, scaled to photons per second, increases by a factor of 2 (Eq.~\ref{NoisePho2SN}), shown as the dashed red line. Because the expectation value of our data lies on top of the theoretical shot-noise limit after the second demodulation stage, shot noise is in fact the dominant noise source in our setup.

This measurement verifies that our system is shot-noise limited and behaves as expected. Because the measurement does not cross the 5-sigma threshold, this also shows that no spurious signals are picked up over the entire 19 day integration time when Laser 2 was turned off.

\subsection{Weak Signal Generation and Detection} 
In order to demonstrate that a signal is observable using heterodyne detection, we generate a beat note between the LO and an ultra-weak sideband of the second laser. We choose a sideband power equivalent to $\sim 10^{-2}$ photons per second. Reducing the signal further was not possible in our current setup as we started to pick up spurious signals electronically. While this has been observed we want to stress that the spurious signal vanishes when the EOM phase modulation is turned off. Thus, it is not an artifact of the second laser field but rather a result of the modulation itself. We assume the issue to be cross-talk between the function generator driving the EOM and the FPGA data acquisition and signal processing card. Further work on generating ultra-weak laser fields without electrical interference is required.

In order to generate a sideband with the specified power, we first remove the ND filters and set the local oscillator to $\bar{P}_{\textrm{LO}}= $ 5 mW and $L_2$ to $\bar{P}_{\textrm{L2}} = $ 6 $\mu$W. Both of these measurements are taken at the photodetector input. The modulation depth is set to $m = 0.0109$ by adjusting the drive amplitude to the EOM. Using Eq.~\ref{SBPower}, the power in the 2nd order sideband (k = +2) is calculated to be on the order of $10^{-15}$ W. The ND filters are placed back into the beam path attenuating the sideband by a factor of $\sim 2 \times 10^5$, yielding $\bar{P}_{\textrm{signal}} = \bar{P}_{\textrm{SB,2}} = 6.33 \times 10^{-21} \, \textrm{W}$. For 1064 nm light, this is equivalent to $3.39 \times 10^{-2}$ photons per second in the sideband we wish to measure.

The CC beat note between $L_1$ and $L_2$ is set to 30 MHz. Phase modulation is done by driving the EOM with a sine wave at 23 MHz + 1.2 Hz. This choice of frequency sets the beat note between the 2nd order sideband and $L_1$ to be at $f_{\textrm{sig}}=$ 16 MHz + 2.4 Hz. With the first demodulation frequency set to $f_1 = 16$ MHz, the beat note of interest is therefore at 2.4 Hz when the data are written to file. These data are then imported into MATLAB where the second demodulation is performed. Finally, we compute $Z_2(N')$ and scale the result to photons per second. 

\begin{figure}[h]
\centering
\includegraphics[width=9cm]{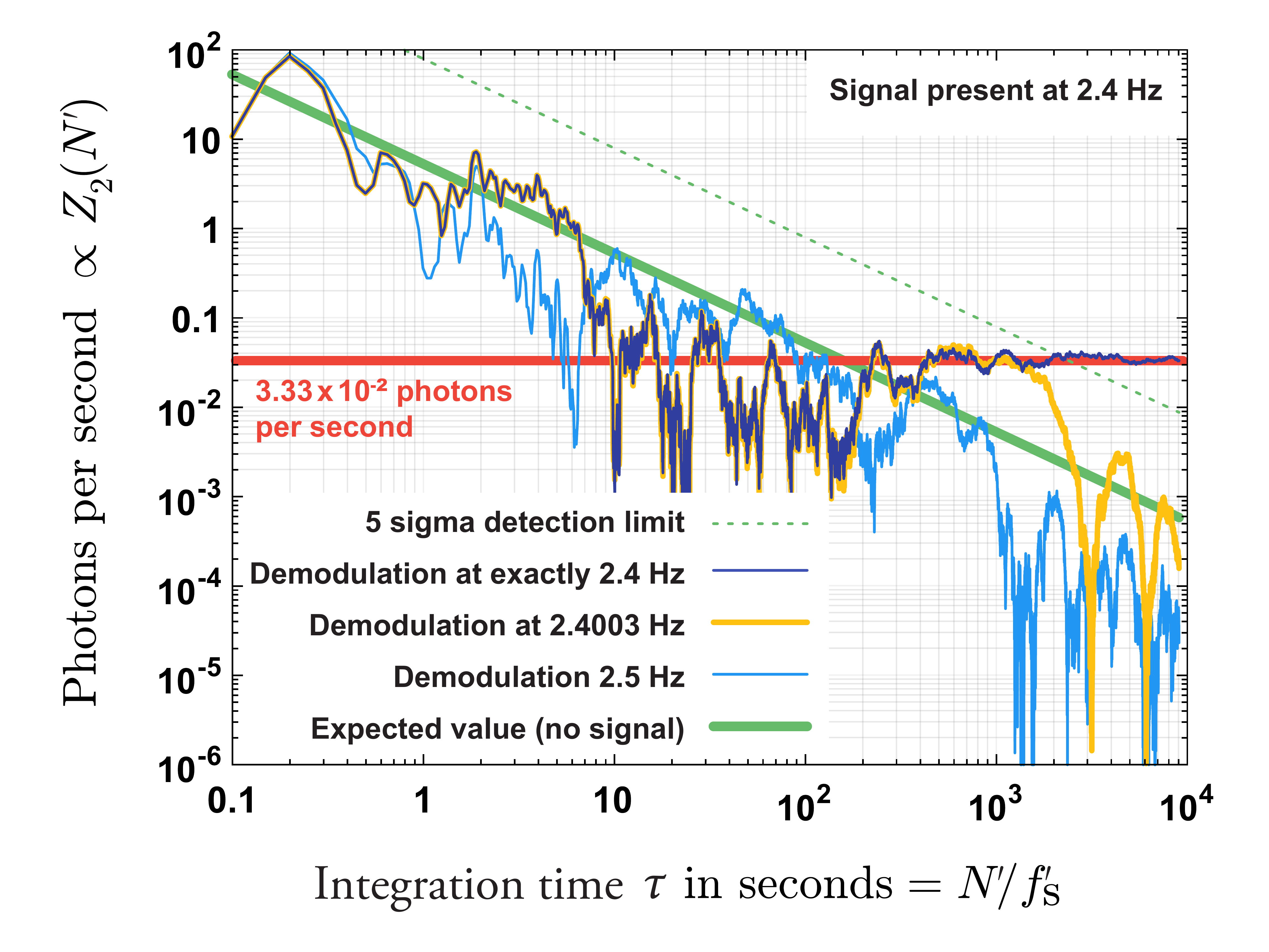}
\caption{Shot-noise limited signal measurement scaled to photons per second. Two demodulation stages are used with a signal present at 2.4 Hz when the data are written to file. When second demodulation is at a frequency $f_2 \neq 2.4$ Hz, the result yields the behavior of noise, shown in light blue. The trend of the expectation value for this level of noise is shown in solid green. The 5-sigma confidence line is shown in dashed green. The result when demodulating at the beat note signal frequency, $f_2 = 2.4$ Hz is shown as the dark blue curve. This curve crosses the 5-sigma line, demonstrating a confident detection. The level that this curve flattens out to yields a rate in the sideband of interest of $3.33 \times 10^{-2}$ photons per second, shown as the red line. Demodulating at a frequency 300 $\mu$Hz away from the beat note signal, shown in yellow, highlights the energy resolution of this detection method.}
\label {fig:DivideByRtN}
\end{figure} 

The results of this measurement are shown in Fig.~\ref{fig:DivideByRtN}. Demodulating at a frequency not equal to any beat note signal frequency demonstrates the expected behavior of noise. This is shown by the light blue curve for which a demodulation 0.1 Hz away from the 2.4 Hz beat note signal was used. In this case, no coherent signal can accumulate and the only influence at the demodulation frequency is noise. This matches the trend of the expectation value of the noise, shown in solid green.   

Demodulating at the beat note signal frequency of $f_\delta = 2.4 \, \textrm{Hz}$, shown as the dark blue curve, initially behaves as noise. The noise dominance continues until the signal begins to take over, causing the curve to flatten out and subsequently cross the 5-sigma threshold. The level at which this curve flattens out yields a rate for the sideband of $3.33 \times 10^{-2}$ photons per second. The measured photon rate differs from expectation by 2\%, a difference that we find acceptable. This error arises from both laser power measurements and modulation depth measurements. The constant level crosses the solid green expected noise curve at $\sim$ 170 seconds, in agreement with the expected $\tau_{\text{x}}$. A 5-sigma confidence detection is made after $\sim$ 2500 seconds of integration time, in agreement with the expected $\tau_{\textrm{5s}}$. We therefore demonstrate that our experimental setup is viable for both generating and detecting sub-photon per second level signals using optical heterodyne interferometry.  

Demodulation 300 $\mu$Hz away from the beat note signal demonstrates the importance of maintaining phase coherence throughout the entire measurement. In this case, shown in yellow, the demodulation waveform drifts in and out of phase with the beat note signal. When this happens, the integrated I and Q values begin to oscillate with the difference frequency, $|f_\delta - f_2|$. This causes $Z_2(N')$ to fall off as a sinc function, preventing it from crossing the 5-sigma threshold.  

\section{Conclusion}
These measurements demonstrate that heterodyne interferometry can be applied as a single photon detector. It however requires that the demodulation waveform maintains phase coherence with the signal during the entire integration time. Measurements at the shot-noise limit with Laser 2 off did not reveal any spurious signals that would degrade the sensitivity of our setup after 19 days of integration. Therefore we can detect coherent signals with field strengths equivalent to about $10^{-5}$ photons per second (1-sigma limit). In order to claim 5-sigma confident detection for such signals we require an integration time of approximately 47 days.

We also demonstrate successful generation and detection of a signal with a field strength on the order of $10^{-2}$ photons per second. Longer integration times and improvements in the generation of ultra-weak laser fields are required to achieve lower power levels which are comparable to the projected sensitivity of ALPS IIc. Work on the generation, implementation, and detection of weaker signal fields is currently ongoing. 

Our results also highlight the importance of maintaining phase coherence and stability throughout the measurement. These limitations to heterodyne detection must be taken into account during implementation into ALPS II. For example, while our measurements are performed using free field propagating beams, ALPS II will make use of two Fabry Perot cavities. The phase noise induced by these cavities must be kept at a low enough level to prevent any notable degradation of the signal.  

While this detection method emerged from the need of a single photon detector for the ALPS II experiment, heterodyne interferometric detection of weak fields can be modified for a variety of applications. Although this technique is demonstrated here using 1064 nm laser light, it can be extended to any wavelength provided that noise and the coherent signal can be decoupled. The versatility of heterodyne detection makes it applicable to a broad range of fields including astronomy, classical communications, and biomedical imaging \cite{Hadfield2009}, as long as the signal field is coherent and its frequency is known.

\section{Acknowledgements}
The authors would like to thank Johannes Eichholz for the phase meter design used for real time signal demodulation. This material is based upon work supported by the National Science Foundation under Grants PHY-1505743 and PHY-1802006 and by the Heising-Simons Foundation under Grant No. 2015-154.

\newpage

\bibliography{bib}

\begin{thebibliography}{35}%
\makeatletter
\providecommand \@ifxundefined [1]{%
 \@ifx{#1\undefined}
}%
\providecommand \@ifnum [1]{%
 \ifnum #1\expandafter \@firstoftwo
 \else \expandafter \@secondoftwo
 \fi
}%
\providecommand \@ifx [1]{%
 \ifx #1\expandafter \@firstoftwo
 \else \expandafter \@secondoftwo
 \fi
}%
\providecommand \natexlab [1]{#1}%
\providecommand \enquote  [1]{``#1''}%
\providecommand \bibnamefont  [1]{#1}%
\providecommand \bibfnamefont [1]{#1}%
\providecommand \citenamefont [1]{#1}%
\providecommand \href@noop [0]{\@secondoftwo}%
\providecommand \href [0]{\begingroup \@sanitize@url \@href}%
\providecommand \@href[1]{\@@startlink{#1}\@@href}%
\providecommand \@@href[1]{\endgroup#1\@@endlink}%
\providecommand \@sanitize@url [0]{\catcode `\\12\catcode `\$12\catcode
  `\&12\catcode `\#12\catcode `\^12\catcode `\_12\catcode `\%12\relax}%
\providecommand \@@startlink[1]{}%
\providecommand \@@endlink[0]{}%
\providecommand \url  [0]{\begingroup\@sanitize@url \@url }%
\providecommand \@url [1]{\endgroup\@href {#1}{\urlprefix }}%
\providecommand \urlprefix  [0]{URL }%
\providecommand \Eprint [0]{\href }%
\providecommand \doibase [0]{http://dx.doi.org/}%
\providecommand \selectlanguage [0]{\@gobble}%
\providecommand \bibinfo  [0]{\@secondoftwo}%
\providecommand \bibfield  [0]{\@secondoftwo}%
\providecommand \translation [1]{[#1]}%
\providecommand \BibitemOpen [0]{}%
\providecommand \bibitemStop [0]{}%
\providecommand \bibitemNoStop [0]{.\EOS\space}%
\providecommand \EOS [0]{\spacefactor3000\relax}%
\providecommand \BibitemShut  [1]{\csname bibitem#1\endcsname}%
\let\auto@bib@innerbib\@empty
\bibitem [{\citenamefont {Baker}\ \emph {et~al.}(2006)\citenamefont {Baker},
  \citenamefont {Doyle}, \citenamefont {Geltenbort}, \citenamefont {Green},
  \citenamefont {van~der Grinten}, \citenamefont {Harris}, \citenamefont
  {Iaydjiev}, \citenamefont {Ivanov}, \citenamefont {May}, \citenamefont
  {Pendlebury}, \citenamefont {Richardson}, \citenamefont {Shiers},\ and\
  \citenamefont {Smith}}]{Neutron}%
  \BibitemOpen
  \bibfield  {author} {\bibinfo {author} {\bibfnamefont {C.~A.}\ \bibnamefont
  {Baker}}, \bibinfo {author} {\bibfnamefont {D.~D.}\ \bibnamefont {Doyle}},
  \bibinfo {author} {\bibfnamefont {P.}~\bibnamefont {Geltenbort}}, \bibinfo
  {author} {\bibfnamefont {K.}~\bibnamefont {Green}}, \bibinfo {author}
  {\bibfnamefont {M.~G.~D.}\ \bibnamefont {van~der Grinten}}, \bibinfo {author}
  {\bibfnamefont {P.~G.}\ \bibnamefont {Harris}}, \bibinfo {author}
  {\bibfnamefont {P.}~\bibnamefont {Iaydjiev}}, \bibinfo {author}
  {\bibfnamefont {S.~N.}\ \bibnamefont {Ivanov}}, \bibinfo {author}
  {\bibfnamefont {D.~J.~R.}\ \bibnamefont {May}}, \bibinfo {author}
  {\bibfnamefont {J.~M.}\ \bibnamefont {Pendlebury}}, \bibinfo {author}
  {\bibfnamefont {J.~D.}\ \bibnamefont {Richardson}}, \bibinfo {author}
  {\bibfnamefont {D.}~\bibnamefont {Shiers}}, \ and\ \bibinfo {author}
  {\bibfnamefont {K.~F.}\ \bibnamefont {Smith}},\ }\href {\doibase
  10.1103/PhysRevLett.97.131801} {\bibfield  {journal} {\bibinfo  {journal}
  {Phys. Rev. Lett.}\ }\textbf {\bibinfo {volume} {97}},\ \bibinfo {pages}
  {131801} (\bibinfo {year} {2006})}\BibitemShut {NoStop}%
\bibitem [{\citenamefont {Peccei}\ and\ \citenamefont
  {Quinn}(1977)}]{PhysRevLett.38.1440}%
  \BibitemOpen
  \bibfield  {author} {\bibinfo {author} {\bibfnamefont {R.~D.}\ \bibnamefont
  {Peccei}}\ and\ \bibinfo {author} {\bibfnamefont {H.~R.}\ \bibnamefont
  {Quinn}},\ }\href {\doibase 10.1103/PhysRevLett.38.1440} {\bibfield
  {journal} {\bibinfo  {journal} {Phys. Rev. Lett.}\ }\textbf {\bibinfo
  {volume} {38}},\ \bibinfo {pages} {1440} (\bibinfo {year}
  {1977})}\BibitemShut {NoStop}%
\bibitem [{\citenamefont {Weinberg}(1978)}]{Weinberg}%
  \BibitemOpen
  \bibfield  {author} {\bibinfo {author} {\bibfnamefont {S.}~\bibnamefont
  {Weinberg}},\ }\href {\doibase 10.1103/PhysRevLett.40.223} {\bibfield
  {journal} {\bibinfo  {journal} {Phys. Rev. Lett.}\ }\textbf {\bibinfo
  {volume} {40}},\ \bibinfo {pages} {223} (\bibinfo {year} {1978})}\BibitemShut
  {NoStop}%
\bibitem [{\citenamefont {Wilczek}(1978)}]{Wilczek}%
  \BibitemOpen
  \bibfield  {author} {\bibinfo {author} {\bibfnamefont {F.}~\bibnamefont
  {Wilczek}},\ }\href {\doibase 10.1103/PhysRevLett.40.279} {\bibfield
  {journal} {\bibinfo  {journal} {Phys. Rev. Lett.}\ }\textbf {\bibinfo
  {volume} {40}},\ \bibinfo {pages} {279} (\bibinfo {year} {1978})}\BibitemShut
  {NoStop}%
\bibitem [{\citenamefont {Olive}\ and\ \citenamefont
  {Group}(2014)}]{ParticleReview}%
  \BibitemOpen
  \bibfield  {author} {\bibinfo {author} {\bibfnamefont {K.}~\bibnamefont
  {Olive}}\ and\ \bibinfo {author} {\bibfnamefont {P.~D.}\ \bibnamefont
  {Group}},\ }\href {http://stacks.iop.org/1674-1137/38/i=9/a=090001}
  {\bibfield  {journal} {\bibinfo  {journal} {Chinese Physics C}\ }\textbf
  {\bibinfo {volume} {38}} (\bibinfo {year} {2014})}\BibitemShut {NoStop}%
\bibitem [{\citenamefont {Svrcek}\ and\ \citenamefont {Witten}(2006)}]{Svrcek}%
  \BibitemOpen
  \bibfield  {author} {\bibinfo {author} {\bibfnamefont {P.}~\bibnamefont
  {Svrcek}}\ and\ \bibinfo {author} {\bibfnamefont {E.}~\bibnamefont
  {Witten}},\ }\href {http://stacks.iop.org/1126-6708/2006/i=06/a=051}
  {\bibfield  {journal} {\bibinfo  {journal} {Journal of High Energy Physics}\
  }\textbf {\bibinfo {volume} {2006}},\ \bibinfo {pages} {051} (\bibinfo {year}
  {2006})}\BibitemShut {NoStop}%
\bibitem [{\citenamefont {Graham}\ \emph {et~al.}(2015)\citenamefont {Graham},
  \citenamefont {Irastorza}, \citenamefont {Lamoreaux}, \citenamefont
  {Lindner},\ and\ \citenamefont {van Bibber}}]{ExperimentalSearch}%
  \BibitemOpen
  \bibfield  {author} {\bibinfo {author} {\bibfnamefont {P.~W.}\ \bibnamefont
  {Graham}}, \bibinfo {author} {\bibfnamefont {I.~G.}\ \bibnamefont
  {Irastorza}}, \bibinfo {author} {\bibfnamefont {S.~K.}\ \bibnamefont
  {Lamoreaux}}, \bibinfo {author} {\bibfnamefont {A.}~\bibnamefont {Lindner}},
  \ and\ \bibinfo {author} {\bibfnamefont {K.~A.}\ \bibnamefont {van Bibber}},\
  }\href {\doibase 10.1146/annurev-nucl-102014-022120} {\bibfield  {journal}
  {\bibinfo  {journal} {Annual Review of Nuclear and Particle Science}\
  }\textbf {\bibinfo {volume} {65}},\ \bibinfo {pages} {485} (\bibinfo {year}
  {2015})}\BibitemShut {NoStop}%
\bibitem [{\citenamefont {Meyer}\ \emph {et~al.}(2013)\citenamefont {Meyer},
  \citenamefont {Horns},\ and\ \citenamefont {Raue}}]{Meyer2013}%
  \BibitemOpen
  \bibfield  {author} {\bibinfo {author} {\bibfnamefont {M.}~\bibnamefont
  {Meyer}}, \bibinfo {author} {\bibfnamefont {D.}~\bibnamefont {Horns}}, \ and\
  \bibinfo {author} {\bibfnamefont {M.}~\bibnamefont {Raue}},\ }\href {\doibase
  10.1103/PhysRevD.87.035027} {\bibfield  {journal} {\bibinfo  {journal} {Phys.
  Rev. D}\ }\textbf {\bibinfo {volume} {87}},\ \bibinfo {pages} {035027}
  (\bibinfo {year} {2013})}\BibitemShut {NoStop}%
\bibitem [{\citenamefont {S.~Catal{\'a}n}\ \emph {et~al.}(2008)\citenamefont
  {S.~Catal{\'a}n}, \citenamefont {Garc{\'i}a-Berro},\ and\ \citenamefont
  {Torres}}]{WhiteDwarf}%
  \BibitemOpen
  \bibfield  {author} {\bibinfo {author} {\bibfnamefont {J.~I.}\ \bibnamefont
  {S.~Catal{\'a}n}}, \bibinfo {author} {\bibfnamefont {E.}~\bibnamefont
  {Garc{\'i}a-Berro}}, \ and\ \bibinfo {author} {\bibfnamefont
  {S.}~\bibnamefont {Torres}},\ }\href
  {http://stacks.iop.org/1538-4357/682/i=2/a=L109} {\bibfield  {journal}
  {\bibinfo  {journal} {The Astrophysical Journal Letters}\ }\textbf {\bibinfo
  {volume} {682}},\ \bibinfo {pages} {L109} (\bibinfo {year}
  {2008})}\BibitemShut {NoStop}%
\bibitem [{\citenamefont {Sikivie}(1983)}]{sikivie1983experimental}%
  \BibitemOpen
  \bibfield  {author} {\bibinfo {author} {\bibfnamefont {P.}~\bibnamefont
  {Sikivie}},\ }\href {\doibase 10.1103/PhysRevLett.51.1415} {\bibfield
  {journal} {\bibinfo  {journal} {Phys. Rev. Lett.}\ }\textbf {\bibinfo
  {volume} {51}},\ \bibinfo {pages} {1415} (\bibinfo {year}
  {1983})}\BibitemShut {NoStop}%
\bibitem [{\citenamefont {Primakoff}(1951)}]{PhysRev.81.899}%
  \BibitemOpen
  \bibfield  {author} {\bibinfo {author} {\bibfnamefont {H.}~\bibnamefont
  {Primakoff}},\ }\href {\doibase 10.1103/PhysRev.81.899} {\bibfield  {journal}
  {\bibinfo  {journal} {Phys. Rev.}\ }\textbf {\bibinfo {volume} {81}},\
  \bibinfo {pages} {899} (\bibinfo {year} {1951})}\BibitemShut {NoStop}%
\bibitem [{\citenamefont {Du}\ \emph {et~al.}(2018)\citenamefont {Du},
  \citenamefont {Force}, \citenamefont {Khatiwada}, \citenamefont {Lentz},
  \citenamefont {Ottens}, \citenamefont {Rosenberg}, \citenamefont {Rybka},
  \citenamefont {Carosi}, \citenamefont {Woollett}, \citenamefont {Bowring},
  \citenamefont {Chou}, \citenamefont {Sonnenschein}, \citenamefont {Wester},
  \citenamefont {Boutan}, \citenamefont {Oblath}, \citenamefont {Bradley},
  \citenamefont {Daw}, \citenamefont {Dixit}, \citenamefont {Clarke},
  \citenamefont {O'Kelley}, \citenamefont {Crisosto}, \citenamefont {Gleason},
  \citenamefont {Jois}, \citenamefont {Sikivie}, \citenamefont {Stern},
  \citenamefont {Sullivan}, \citenamefont {Tanner},\ and\ \citenamefont
  {Hilton}}]{ADMXSuggested}%
  \BibitemOpen
  \bibfield  {author} {\bibinfo {author} {\bibfnamefont {N.}~\bibnamefont
  {Du}}, \bibinfo {author} {\bibfnamefont {N.}~\bibnamefont {Force}}, \bibinfo
  {author} {\bibfnamefont {R.}~\bibnamefont {Khatiwada}}, \bibinfo {author}
  {\bibfnamefont {E.}~\bibnamefont {Lentz}}, \bibinfo {author} {\bibfnamefont
  {R.}~\bibnamefont {Ottens}}, \bibinfo {author} {\bibfnamefont {L.~J.}\
  \bibnamefont {Rosenberg}}, \bibinfo {author} {\bibfnamefont {G.}~\bibnamefont
  {Rybka}}, \bibinfo {author} {\bibfnamefont {G.}~\bibnamefont {Carosi}},
  \bibinfo {author} {\bibfnamefont {N.}~\bibnamefont {Woollett}}, \bibinfo
  {author} {\bibfnamefont {D.}~\bibnamefont {Bowring}}, \bibinfo {author}
  {\bibfnamefont {A.~S.}\ \bibnamefont {Chou}}, \bibinfo {author}
  {\bibfnamefont {A.}~\bibnamefont {Sonnenschein}}, \bibinfo {author}
  {\bibfnamefont {W.}~\bibnamefont {Wester}}, \bibinfo {author} {\bibfnamefont
  {C.}~\bibnamefont {Boutan}}, \bibinfo {author} {\bibfnamefont {N.~S.}\
  \bibnamefont {Oblath}}, \bibinfo {author} {\bibfnamefont {R.}~\bibnamefont
  {Bradley}}, \bibinfo {author} {\bibfnamefont {E.~J.}\ \bibnamefont {Daw}},
  \bibinfo {author} {\bibfnamefont {A.~V.}\ \bibnamefont {Dixit}}, \bibinfo
  {author} {\bibfnamefont {J.}~\bibnamefont {Clarke}}, \bibinfo {author}
  {\bibfnamefont {S.~R.}\ \bibnamefont {O'Kelley}}, \bibinfo {author}
  {\bibfnamefont {N.}~\bibnamefont {Crisosto}}, \bibinfo {author}
  {\bibfnamefont {J.~R.}\ \bibnamefont {Gleason}}, \bibinfo {author}
  {\bibfnamefont {S.}~\bibnamefont {Jois}}, \bibinfo {author} {\bibfnamefont
  {P.}~\bibnamefont {Sikivie}}, \bibinfo {author} {\bibfnamefont
  {I.}~\bibnamefont {Stern}}, \bibinfo {author} {\bibfnamefont {N.~S.}\
  \bibnamefont {Sullivan}}, \bibinfo {author} {\bibfnamefont {D.~B.}\
  \bibnamefont {Tanner}}, \ and\ \bibinfo {author} {\bibfnamefont {G.~C.}\
  \bibnamefont {Hilton}} (\bibinfo {collaboration} {ADMX Collaboration}),\
  }\href {\doibase 10.1103/PhysRevLett.120.151301} {\bibfield  {journal}
  {\bibinfo  {journal} {Phys. Rev. Lett.}\ }\textbf {\bibinfo {volume} {120}},\
  \bibinfo {pages} {151301} (\bibinfo {year} {2018})}\BibitemShut {NoStop}%
\bibitem [{\citenamefont {Zioutas~et. al}(2005)}]{PhysRevLett.94.121301}%
  \BibitemOpen
  \bibfield  {author} {\bibinfo {author} {\bibfnamefont {K.}~\bibnamefont
  {Zioutas~et. al}} (\bibinfo {collaboration} {CAST Collaboration}),\ }\href
  {\doibase 10.1103/PhysRevLett.94.121301} {\bibfield  {journal} {\bibinfo
  {journal} {Phys. Rev. Lett.}\ }\textbf {\bibinfo {volume} {94}},\ \bibinfo
  {pages} {121301} (\bibinfo {year} {2005})}\BibitemShut {NoStop}%
\bibitem [{\citenamefont {Van~Bibber}\ \emph {et~al.}(1987)\citenamefont
  {Van~Bibber}, \citenamefont {Dagdeviren}, \citenamefont {Koonin},
  \citenamefont {Kerman},\ and\ \citenamefont {Nelson}}]{Bibber}%
  \BibitemOpen
  \bibfield  {author} {\bibinfo {author} {\bibfnamefont {K.}~\bibnamefont
  {Van~Bibber}}, \bibinfo {author} {\bibfnamefont {N.~R.}\ \bibnamefont
  {Dagdeviren}}, \bibinfo {author} {\bibfnamefont {S.~E.}\ \bibnamefont
  {Koonin}}, \bibinfo {author} {\bibfnamefont {A.~K.}\ \bibnamefont {Kerman}},
  \ and\ \bibinfo {author} {\bibfnamefont {H.~N.}\ \bibnamefont {Nelson}},\
  }\href {\doibase 10.1103/PhysRevLett.59.759} {\bibfield  {journal} {\bibinfo
  {journal} {Phys. Rev. Lett.}\ }\textbf {\bibinfo {volume} {59}},\ \bibinfo
  {pages} {759} (\bibinfo {year} {1987})}\BibitemShut {NoStop}%
\bibitem [{\citenamefont {Robilliard}\ \emph {et~al.}(2007)\citenamefont
  {Robilliard}, \citenamefont {Battesti}, \citenamefont {Fouch\'e},
  \citenamefont {Mauchain}, \citenamefont {Sautivet}, \citenamefont
  {Amiranoff},\ and\ \citenamefont {Rizzo}}]{Robilliard}%
  \BibitemOpen
  \bibfield  {author} {\bibinfo {author} {\bibfnamefont {C.}~\bibnamefont
  {Robilliard}}, \bibinfo {author} {\bibfnamefont {R.}~\bibnamefont
  {Battesti}}, \bibinfo {author} {\bibfnamefont {M.}~\bibnamefont {Fouch\'e}},
  \bibinfo {author} {\bibfnamefont {J.}~\bibnamefont {Mauchain}}, \bibinfo
  {author} {\bibfnamefont {A.-M.}\ \bibnamefont {Sautivet}}, \bibinfo {author}
  {\bibfnamefont {F.}~\bibnamefont {Amiranoff}}, \ and\ \bibinfo {author}
  {\bibfnamefont {C.}~\bibnamefont {Rizzo}},\ }\href {\doibase
  10.1103/PhysRevLett.99.190403} {\bibfield  {journal} {\bibinfo  {journal}
  {Phys. Rev. Lett.}\ }\textbf {\bibinfo {volume} {99}},\ \bibinfo {pages}
  {190403} (\bibinfo {year} {2007})}\BibitemShut {NoStop}%
\bibitem [{\citenamefont {Chou}\ \emph {et~al.}(2008)\citenamefont {Chou},
  \citenamefont {Wester}, \citenamefont {Baumbaugh}, \citenamefont {Gustafson},
  \citenamefont {Irizarry-Valle}, \citenamefont {Mazur}, \citenamefont
  {Steffen}, \citenamefont {Tomlin}, \citenamefont {Yang},\ and\ \citenamefont
  {Yoo}}]{GammeVSuggested}%
  \BibitemOpen
  \bibfield  {author} {\bibinfo {author} {\bibfnamefont {A.~S.}\ \bibnamefont
  {Chou}}, \bibinfo {author} {\bibfnamefont {W.}~\bibnamefont {Wester}},
  \bibinfo {author} {\bibfnamefont {A.}~\bibnamefont {Baumbaugh}}, \bibinfo
  {author} {\bibfnamefont {H.~R.}\ \bibnamefont {Gustafson}}, \bibinfo {author}
  {\bibfnamefont {Y.}~\bibnamefont {Irizarry-Valle}}, \bibinfo {author}
  {\bibfnamefont {P.~O.}\ \bibnamefont {Mazur}}, \bibinfo {author}
  {\bibfnamefont {J.~H.}\ \bibnamefont {Steffen}}, \bibinfo {author}
  {\bibfnamefont {R.}~\bibnamefont {Tomlin}}, \bibinfo {author} {\bibfnamefont
  {X.}~\bibnamefont {Yang}}, \ and\ \bibinfo {author} {\bibfnamefont
  {J.}~\bibnamefont {Yoo}},\ }\href {\doibase 10.1103/PhysRevLett.100.080402}
  {\bibfield  {journal} {\bibinfo  {journal} {Phys. Rev. Lett.}\ }\textbf
  {\bibinfo {volume} {100}},\ \bibinfo {pages} {080402} (\bibinfo {year}
  {2008})}\BibitemShut {NoStop}%
\bibitem [{\citenamefont {Afanasev}\ \emph {et~al.}(2008)\citenamefont
  {Afanasev}, \citenamefont {Baker}, \citenamefont {Beard}, \citenamefont
  {Biallas}, \citenamefont {Boyce}, \citenamefont {Minarni}, \citenamefont
  {Ramdon}, \citenamefont {Shinn},\ and\ \citenamefont
  {Slocum}}]{LIPPSSuggested}%
  \BibitemOpen
  \bibfield  {author} {\bibinfo {author} {\bibfnamefont {A.}~\bibnamefont
  {Afanasev}}, \bibinfo {author} {\bibfnamefont {O.~K.}\ \bibnamefont {Baker}},
  \bibinfo {author} {\bibfnamefont {K.~B.}\ \bibnamefont {Beard}}, \bibinfo
  {author} {\bibfnamefont {G.}~\bibnamefont {Biallas}}, \bibinfo {author}
  {\bibfnamefont {J.}~\bibnamefont {Boyce}}, \bibinfo {author} {\bibfnamefont
  {M.}~\bibnamefont {Minarni}}, \bibinfo {author} {\bibfnamefont
  {R.}~\bibnamefont {Ramdon}}, \bibinfo {author} {\bibfnamefont
  {M.}~\bibnamefont {Shinn}}, \ and\ \bibinfo {author} {\bibfnamefont
  {P.}~\bibnamefont {Slocum}},\ }\href {\doibase
  10.1103/PhysRevLett.101.120401} {\bibfield  {journal} {\bibinfo  {journal}
  {Phys. Rev. Lett.}\ }\textbf {\bibinfo {volume} {101}},\ \bibinfo {pages}
  {120401} (\bibinfo {year} {2008})}\BibitemShut {NoStop}%
\bibitem [{\citenamefont {Ehret}\ \emph {et~al.}(2010)\citenamefont {Ehret},
  \citenamefont {Frede}, \citenamefont {Ghazaryan}, \citenamefont
  {Hildebrandt}, \citenamefont {Knabbe}, \citenamefont {Kracht}, \citenamefont
  {Lindner}, \citenamefont {List}, \citenamefont {Meier}, \citenamefont
  {Meyer}, \citenamefont {Notz}, \citenamefont {Redondo}, \citenamefont
  {Ringwald}, \citenamefont {Wiedemann},\ and\ \citenamefont
  {Willke}}]{NewALPSResult}%
  \BibitemOpen
  \bibfield  {author} {\bibinfo {author} {\bibfnamefont {K.}~\bibnamefont
  {Ehret}}, \bibinfo {author} {\bibfnamefont {M.}~\bibnamefont {Frede}},
  \bibinfo {author} {\bibfnamefont {S.}~\bibnamefont {Ghazaryan}}, \bibinfo
  {author} {\bibfnamefont {M.}~\bibnamefont {Hildebrandt}}, \bibinfo {author}
  {\bibfnamefont {E.-A.}\ \bibnamefont {Knabbe}}, \bibinfo {author}
  {\bibfnamefont {D.}~\bibnamefont {Kracht}}, \bibinfo {author} {\bibfnamefont
  {A.}~\bibnamefont {Lindner}}, \bibinfo {author} {\bibfnamefont
  {J.}~\bibnamefont {List}}, \bibinfo {author} {\bibfnamefont {T.}~\bibnamefont
  {Meier}}, \bibinfo {author} {\bibfnamefont {N.}~\bibnamefont {Meyer}},
  \bibinfo {author} {\bibfnamefont {D.}~\bibnamefont {Notz}}, \bibinfo {author}
  {\bibfnamefont {J.}~\bibnamefont {Redondo}}, \bibinfo {author} {\bibfnamefont
  {A.}~\bibnamefont {Ringwald}}, \bibinfo {author} {\bibfnamefont
  {G.}~\bibnamefont {Wiedemann}}, \ and\ \bibinfo {author} {\bibfnamefont
  {B.}~\bibnamefont {Willke}},\ }\href {\doibase
  https://doi.org/10.1016/j.physletb.2010.04.066} {\bibfield  {journal}
  {\bibinfo  {journal} {Physics Letters B}\ }\textbf {\bibinfo {volume}
  {689}},\ \bibinfo {pages} {149 } (\bibinfo {year} {2010})}\BibitemShut
  {NoStop}%
\bibitem [{\citenamefont {Pugnat}\ \emph {et~al.}(2014)\citenamefont {Pugnat},
  \citenamefont {Ballou}, \citenamefont {Schott}, \citenamefont {Husek},
  \citenamefont {Sulc}, \citenamefont {Deferne}, \citenamefont {Duvillaret},
  \citenamefont {Finger}, \citenamefont {Finger}, \citenamefont {Flekova},
  \citenamefont {Hosek}, \citenamefont {Jary}, \citenamefont {Jost},
  \citenamefont {Kral}, \citenamefont {Kunc}, \citenamefont {Macuchova},
  \citenamefont {Meissner}, \citenamefont {Morville}, \citenamefont {Romanini},
  \citenamefont {Siemko}, \citenamefont {Slunecka}, \citenamefont {Vitrant},\
  and\ \citenamefont {Zicha}}]{OSQARSuggested}%
  \BibitemOpen
  \bibfield  {author} {\bibinfo {author} {\bibfnamefont {P.}~\bibnamefont
  {Pugnat}}, \bibinfo {author} {\bibfnamefont {R.}~\bibnamefont {Ballou}},
  \bibinfo {author} {\bibfnamefont {M.}~\bibnamefont {Schott}}, \bibinfo
  {author} {\bibfnamefont {T.}~\bibnamefont {Husek}}, \bibinfo {author}
  {\bibfnamefont {M.}~\bibnamefont {Sulc}}, \bibinfo {author} {\bibfnamefont
  {G.}~\bibnamefont {Deferne}}, \bibinfo {author} {\bibfnamefont
  {L.}~\bibnamefont {Duvillaret}}, \bibinfo {author} {\bibfnamefont
  {M.}~\bibnamefont {Finger}}, \bibinfo {author} {\bibfnamefont
  {M.}~\bibnamefont {Finger}}, \bibinfo {author} {\bibfnamefont
  {L.}~\bibnamefont {Flekova}}, \bibinfo {author} {\bibfnamefont
  {J.}~\bibnamefont {Hosek}}, \bibinfo {author} {\bibfnamefont
  {V.}~\bibnamefont {Jary}}, \bibinfo {author} {\bibfnamefont {R.}~\bibnamefont
  {Jost}}, \bibinfo {author} {\bibfnamefont {M.}~\bibnamefont {Kral}}, \bibinfo
  {author} {\bibfnamefont {S.}~\bibnamefont {Kunc}}, \bibinfo {author}
  {\bibfnamefont {K.}~\bibnamefont {Macuchova}}, \bibinfo {author}
  {\bibfnamefont {K.~A.}\ \bibnamefont {Meissner}}, \bibinfo {author}
  {\bibfnamefont {J.}~\bibnamefont {Morville}}, \bibinfo {author}
  {\bibfnamefont {D.}~\bibnamefont {Romanini}}, \bibinfo {author}
  {\bibfnamefont {A.}~\bibnamefont {Siemko}}, \bibinfo {author} {\bibfnamefont
  {M.}~\bibnamefont {Slunecka}}, \bibinfo {author} {\bibfnamefont
  {G.}~\bibnamefont {Vitrant}}, \ and\ \bibinfo {author} {\bibfnamefont
  {J.}~\bibnamefont {Zicha}},\ }\href {\doibase 10.1140/epjc/s10052-014-3027-8}
  {\bibfield  {journal} {\bibinfo  {journal} {The European Physical Journal C}\
  }\textbf {\bibinfo {volume} {74}},\ \bibinfo {pages} {3027} (\bibinfo {year}
  {2014})}\BibitemShut {NoStop}%
\bibitem [{\citenamefont {Hoogeveen}\ and\ \citenamefont
  {Ziegenhagen}(1991)}]{HOOGEVEEN19913}%
  \BibitemOpen
  \bibfield  {author} {\bibinfo {author} {\bibfnamefont {F.}~\bibnamefont
  {Hoogeveen}}\ and\ \bibinfo {author} {\bibfnamefont {T.}~\bibnamefont
  {Ziegenhagen}},\ }\href {\doibase
  https://doi.org/10.1016/0550-3213(91)90528-6} {\bibfield  {journal} {\bibinfo
   {journal} {Nuclear Physics B}\ }\textbf {\bibinfo {volume} {{358}}},\
  \bibinfo {pages} {3 } (\bibinfo {year} {1991})}\BibitemShut {NoStop}%
\bibitem [{\citenamefont {{Y. Fukuda, T. Kohmoto, S. Nakajima, and M.
  Kunitomo}}(1996)}]{FUKUDA1996363}%
  \BibitemOpen
  \bibfield  {author} {\bibinfo {author} {\bibnamefont {{Y. Fukuda, T. Kohmoto,
  S. Nakajima, and M. Kunitomo}}},\ }\href {\doibase
  https://doi.org/10.1016/0960-8974(96)83672-2} {\bibfield  {journal} {\bibinfo
   {journal} {Progress in Crystal Growth and Characterization of Materials}\
  }\textbf {\bibinfo {volume} {{33}}},\ \bibinfo {pages} {363 } (\bibinfo
  {year} {1996})}\BibitemShut {NoStop}%
\bibitem [{\citenamefont {Sikivie}\ \emph {et~al.}(2007)\citenamefont
  {Sikivie}, \citenamefont {Tanner},\ and\ \citenamefont {van
  Bibber}}]{SikivieEnhanced}%
  \BibitemOpen
  \bibfield  {author} {\bibinfo {author} {\bibfnamefont {P.}~\bibnamefont
  {Sikivie}}, \bibinfo {author} {\bibfnamefont {D.~B.}\ \bibnamefont {Tanner}},
  \ and\ \bibinfo {author} {\bibfnamefont {K.}~\bibnamefont {van Bibber}},\
  }\href {\doibase 10.1103/PhysRevLett.98.172002} {\bibfield  {journal}
  {\bibinfo  {journal} {Phys. Rev. Lett.}\ }\textbf {\bibinfo {volume} {98}},\
  \bibinfo {pages} {172002} (\bibinfo {year} {2007})}\BibitemShut {NoStop}%
\bibitem [{\citenamefont {Mueller}\ \emph {et~al.}(2009)\citenamefont
  {Mueller}, \citenamefont {Sikivie}, \citenamefont {Tanner},\ and\
  \citenamefont {van Bibber}}]{Mueller2009}%
  \BibitemOpen
  \bibfield  {author} {\bibinfo {author} {\bibfnamefont {G.}~\bibnamefont
  {Mueller}}, \bibinfo {author} {\bibfnamefont {P.}~\bibnamefont {Sikivie}},
  \bibinfo {author} {\bibfnamefont {D.~B.}\ \bibnamefont {Tanner}}, \ and\
  \bibinfo {author} {\bibfnamefont {K.}~\bibnamefont {van Bibber}},\ }\href
  {\doibase 10.1103/PhysRevD.80.072004} {\bibfield  {journal} {\bibinfo
  {journal} {Phys. Rev. D}\ }\textbf {\bibinfo {volume} {80}},\ \bibinfo
  {pages} {072004} (\bibinfo {year} {2009})}\BibitemShut {NoStop}%
\bibitem [{\citenamefont {B\"{a}hre}\ \emph {et~al.}(2013)\citenamefont
  {B\"{a}hre}, \citenamefont {D\"{o}brich}, \citenamefont
  {Dreyling-Eschweiler}, \citenamefont {Ghazaryan}, \citenamefont {Hodajerdi},
  \citenamefont {Horns}, \citenamefont {Januschek}, \citenamefont {Knabbe},
  \citenamefont {Lindner}, \citenamefont {Notz}, \citenamefont {Ringwald},
  \citenamefont {von Seggern}, \citenamefont {Stromhagen}, \citenamefont
  {Trines},\ and\ \citenamefont {Willke}}]{Bahre2013}%
  \BibitemOpen
  \bibfield  {author} {\bibinfo {author} {\bibfnamefont {R.}~\bibnamefont
  {B\"{a}hre}}, \bibinfo {author} {\bibfnamefont {B.}~\bibnamefont
  {D\"{o}brich}}, \bibinfo {author} {\bibfnamefont {J.}~\bibnamefont
  {Dreyling-Eschweiler}}, \bibinfo {author} {\bibfnamefont {S.}~\bibnamefont
  {Ghazaryan}}, \bibinfo {author} {\bibfnamefont {R.}~\bibnamefont
  {Hodajerdi}}, \bibinfo {author} {\bibfnamefont {D.}~\bibnamefont {Horns}},
  \bibinfo {author} {\bibfnamefont {F.}~\bibnamefont {Januschek}}, \bibinfo
  {author} {\bibfnamefont {E.~A.}\ \bibnamefont {Knabbe}}, \bibinfo {author}
  {\bibfnamefont {A.}~\bibnamefont {Lindner}}, \bibinfo {author} {\bibfnamefont
  {D.}~\bibnamefont {Notz}}, \bibinfo {author} {\bibfnamefont {A.}~\bibnamefont
  {Ringwald}}, \bibinfo {author} {\bibfnamefont {J.~E.}\ \bibnamefont {von
  Seggern}}, \bibinfo {author} {\bibfnamefont {R.}~\bibnamefont {Stromhagen}},
  \bibinfo {author} {\bibfnamefont {D.}~\bibnamefont {Trines}}, \ and\ \bibinfo
  {author} {\bibfnamefont {B.}~\bibnamefont {Willke}},\ }\href
  {http://stacks.iop.org/1748-0221/8/i=09/a=T09001} {\bibfield  {journal}
  {\bibinfo  {journal} {Journal of Instrumentation}\ }\textbf {\bibinfo
  {volume} {8}},\ \bibinfo {pages} {T09001} (\bibinfo {year}
  {2013})}\BibitemShut {NoStop}%
\bibitem [{\citenamefont {{A. D. Spector, J. H. P\~{o}ld, R. B\"{a}hre, A.
  Lindner, and B. Willke}}(2016)}]{Spector16}%
  \BibitemOpen
  \bibfield  {author} {\bibinfo {author} {\bibnamefont {{A. D. Spector, J. H.
  P\~{o}ld, R. B\"{a}hre, A. Lindner, and B. Willke}}},\ }\href {\doibase
  10.1364/OE.24.029237} {\bibfield  {journal} {\bibinfo  {journal} {Opt.
  Express}\ }\textbf {\bibinfo {volume} {24}},\ \bibinfo {pages} {29237}
  (\bibinfo {year} {2016})}\BibitemShut {NoStop}%
\bibitem [{\citenamefont {Dreyling-Eschweiler}\ \emph
  {et~al.}(2015)\citenamefont {Dreyling-Eschweiler}, \citenamefont {Bastidon},
  \citenamefont {D\"{o}brich}, \citenamefont {Horns}, \citenamefont
  {Januschek},\ and\ \citenamefont {Lindner}}]{TES}%
  \BibitemOpen
  \bibfield  {author} {\bibinfo {author} {\bibfnamefont {J.}~\bibnamefont
  {Dreyling-Eschweiler}}, \bibinfo {author} {\bibfnamefont {N.}~\bibnamefont
  {Bastidon}}, \bibinfo {author} {\bibfnamefont {B.}~\bibnamefont
  {D\"{o}brich}}, \bibinfo {author} {\bibfnamefont {D.}~\bibnamefont {Horns}},
  \bibinfo {author} {\bibfnamefont {F.}~\bibnamefont {Januschek}}, \ and\
  \bibinfo {author} {\bibfnamefont {A.}~\bibnamefont {Lindner}},\ }\href
  {\doibase 10.1080/09500340.2015.1021723} {\bibfield  {journal} {\bibinfo
  {journal} {Journal of Modern Optics}\ }\textbf {\bibinfo {volume} {62}},\
  \bibinfo {pages} {1132} (\bibinfo {year} {2015})}\BibitemShut {NoStop}%
\bibitem [{\citenamefont {Stoica}\ and\ \citenamefont
  {Moses}(2005)}]{stoica2005spectral}%
  \BibitemOpen
  \bibfield  {author} {\bibinfo {author} {\bibfnamefont {P.}~\bibnamefont
  {Stoica}}\ and\ \bibinfo {author} {\bibfnamefont {R.~L.}\ \bibnamefont
  {Moses}},\ }\href@noop {} {\emph {\bibinfo {title} {Spectral analysis of
  signals}}}\ (\bibinfo  {publisher} {Pearson/Prentice Hall},\ \bibinfo {year}
  {2005})\BibitemShut {NoStop}%
\bibitem [{\citenamefont {Peligrad}\ and\ \citenamefont
  {Wu}(2010)}]{peligrad2010}%
  \BibitemOpen
  \bibfield  {author} {\bibinfo {author} {\bibfnamefont {M.}~\bibnamefont
  {Peligrad}}\ and\ \bibinfo {author} {\bibfnamefont {W.~B.}\ \bibnamefont
  {Wu}},\ }\href {\doibase 10.1214/10-AOP530} {\bibfield  {journal} {\bibinfo
  {journal} {Ann. Probab.}\ }\textbf {\bibinfo {volume} {38}},\ \bibinfo
  {pages} {2009} (\bibinfo {year} {2010})}\BibitemShut {NoStop}%
\bibitem [{\citenamefont {Papoulis}\ and\ \citenamefont
  {Pillai}(2002)}]{papoulis2002probability}%
  \BibitemOpen
  \bibfield  {author} {\bibinfo {author} {\bibfnamefont {A.}~\bibnamefont
  {Papoulis}}\ and\ \bibinfo {author} {\bibfnamefont {S.}~\bibnamefont
  {Pillai}},\ }\href {https://books.google.com/books?id=YYwQAQAAIAAJ} {\emph
  {\bibinfo {title} {{Probability, random variables, and stochastic
  processes}}}},\ McGraw-Hill electrical and electronic engineering series\
  (\bibinfo  {publisher} {McGraw-Hill},\ \bibinfo {year} {2002})\BibitemShut
  {NoStop}%
\bibitem [{\citenamefont {Fox}(2006)}]{fox}%
  \BibitemOpen
  \bibfield  {author} {\bibinfo {author} {\bibfnamefont {M.}~\bibnamefont
  {Fox}},\ }\href@noop {} {\emph {\bibinfo {title} {{Quantum Optics: An
  Introduction (Oxford master series in physics ; 6)}}}}\ (\bibinfo
  {publisher} {Oxford University Press},\ \bibinfo {year} {2006})\BibitemShut
  {NoStop}%
\bibitem [{\citenamefont {Mayer}\ \emph {et~al.}(2003)\citenamefont {Mayer},
  \citenamefont {Rana},\ and\ \citenamefont {Ram}}]{shotnoiseEq}%
  \BibitemOpen
  \bibfield  {author} {\bibinfo {author} {\bibfnamefont {P.~M.}\ \bibnamefont
  {Mayer}}, \bibinfo {author} {\bibfnamefont {F.}~\bibnamefont {Rana}}, \ and\
  \bibinfo {author} {\bibfnamefont {R.~J.}\ \bibnamefont {Ram}},\ }\href
  {\doibase 10.1063/1.1539548} {\bibfield  {journal} {\bibinfo  {journal}
  {Applied Physics Letters}\ }\textbf {\bibinfo {volume} {82}},\ \bibinfo
  {pages} {689} (\bibinfo {year} {2003})}\BibitemShut {NoStop}%
\bibitem [{\citenamefont {Hogenauer}(1981)}]{Hogenauer1981}%
  \BibitemOpen
  \bibfield  {author} {\bibinfo {author} {\bibfnamefont {E.}~\bibnamefont
  {Hogenauer}},\ }\href {\doibase 10.1109/TASSP.1981.1163535} {\bibfield
  {journal} {\bibinfo  {journal} {IEEE Trans. Acoust., Speech, Signal
  Processing}\ }\textbf {\bibinfo {volume} {29}},\ \bibinfo {pages} {155}
  (\bibinfo {year} {1981})}\BibitemShut {NoStop}%
\bibitem [{\citenamefont {Ragaini}\ and\ \citenamefont {Woodman}()}]{RDS}%
  \BibitemOpen
  \bibfield  {author} {\bibinfo {author} {\bibfnamefont {E.}~\bibnamefont
  {Ragaini}}\ and\ \bibinfo {author} {\bibfnamefont {R.~F.}\ \bibnamefont
  {Woodman}},\ }\href {\doibase 10.1029/96RS03580} {\bibfield  {journal}
  {\bibinfo  {journal} {Radio Science}\ }\textbf {\bibinfo {volume} {32}},\
  \bibinfo {pages} {783}}\BibitemShut {NoStop}%
\bibitem [{\citenamefont {Saleh}\ and\ \citenamefont
  {Teich}(2013)}]{saleh2013fundamentals}%
  \BibitemOpen
  \bibfield  {author} {\bibinfo {author} {\bibfnamefont {B.}~\bibnamefont
  {Saleh}}\ and\ \bibinfo {author} {\bibfnamefont {M.}~\bibnamefont {Teich}},\
  }\href {https://books.google.com/books?id=Qfeosgu08u8C} {\emph {\bibinfo
  {title} {Fundamentals of Photonics}}},\ Wiley Series in Pure and Applied
  Optics\ (\bibinfo  {publisher} {Wiley},\ \bibinfo {year} {2013})\BibitemShut
  {NoStop}%
\bibitem [{\citenamefont {Hadfield}(2009)}]{Hadfield2009}%
  \BibitemOpen
  \bibfield  {author} {\bibinfo {author} {\bibfnamefont {R.~H.}\ \bibnamefont
  {Hadfield}},\ }\href {http://dx.doi.org/10.1038/nphoton.2009.230} {\bibfield
  {journal} {\bibinfo  {journal} {Nature Photonics}\ }\textbf {\bibinfo
  {volume} {3}},\ \bibinfo {pages} {696} (\bibinfo {year} {2009})}\BibitemShut
  {NoStop}%
\end{thebibliography}%
\bibliographystyle{apsrev4-1}

\end{document}